\documentclass[sigconf,nonacm]{acmart}

\usepackage[export]{adjustbox}

\usepackage{amsmath}
\usepackage{balance}
\usepackage{booktabs}
\usepackage{comment}
\usepackage{pifont}
\usepackage{placeins}
\usepackage{siunitx}
\usepackage{textcomp}
\usepackage{tikz}
\usepackage{xspace}

\newcommand{\shortX}{\textsf{X}\xspace}
\newcommand{\shortY}{\textsf{Y}\xspace}

\newcommand{\shortA}{\textsf{A}\xspace}
\newcommand{\shortB}{\textsf{B}\xspace}
\newcommand{\shortC}{\textsf{C}\xspace}
\newcommand{\shortD}{\textsf{D}\xspace}

\newcommand{\fullX}{Vehicle \shortX}
\newcommand{\fullY}{Vehicle \shortY}

\newcommand{\fullA}{Charger \shortA}
\newcommand{\fullB}{Charger \shortB}
\newcommand{\fullC}{Charger \shortC}
\newcommand{\fullD}{Charger \shortD}

\newcommand{\currentdemandreq}{\texttt{Current\-Demand\-Req}\xspace}
\newcommand{\currentdemandres}{\texttt{Current\-Demand\-Res}\xspace}

\tikzset{pythondash/.style={thick,dash pattern=on 2pt off 1pt}}

\DeclareRobustCommand{\legendline}[2]{%
  \tikz[baseline=-0.6ex] \draw[#1, line width=1pt, #2] (0,0) -- (8pt,0);%
}
\definecolor{cyclerblue}{RGB}{0,119,187}
\definecolor{cyclerorange}{RGB}{238,119,51}
\definecolor{cyclergreen}{RGB}{0,153,136}
\definecolor{cyclerpink}{RGB}{238,51,119}

\begin{document}

\title{DCeption: Real-world Wireless Man-in-the-Middle\\Attacks Against CCS EV Charging}
\renewcommand{\shorttitle}{DCeption: Real-world Wireless MitM Attacks Against CCS EV Charging}

\author{Marcell Szakály}
\affiliation{%
  \institution{University of Oxford}
  \city{Oxford}
  \country{United Kingdom}
}
\author{Martin Strohmeier}
\affiliation{%
  \institution{armasuisse S+T}
  \country{Switzerland}
}

\author{Ivan Martinovic}
\affiliation{%
  \institution{University of Oxford}
  \city{Oxford}
  \country{United Kingdom}
}

\author{Sebastian Köhler}
\affiliation{%
  \institution{University of Oxford}
  \city{Oxford}
 \country{United Kingdom}
}

\renewcommand{\shortauthors}{Szakály et al.}

\begin{abstract}

The adoption of Electric Vehicles (EVs) is happening at a rapid pace.
To ensure fast and safe charging, complex communication is required between the vehicle and the charging station.
In the globally used Combined Charging System (CCS), this communication is carried over the HomePlug Green PHY (HPGP) physical layer.
However, HPGP is known to suffer from wireless leakage, which may expose this data link to nearby attackers.

In this paper, we examine active wireless attacks against CCS, and study the impact they can have.
We present the first real-time Software-Defined Radio (SDR) implementation of HPGP, granting unprecedented access to the communications within the charging cables.
We analyze the characteristics of 2,750 real-world charging sessions to understand the timing constraints for hijacking. Using novel techniques to increase the attacks' reliability, we design a robust wireless Man-in-the-Middle evaluation framework for CCS.

We demonstrate full control over TLS usage and CCS protocol version negotiation, including TLS stripping attacks.
We investigate how real devices respond to safety-critical MitM attacks, which modify power delivery information, and found target vehicles to be highly permissive.
First, we caused a vehicle to display charging power exceeding \SI{900}{\kilo \watt} on the dashboard, while receiving only \SI{40}{\kilo \watt}.
Second, we remotely overcharged a vehicle, at twice the requested current for \SI{17}{\second} before the vehicle triggered the emergency shutdown.
Finally, we propose a backwards-compatible, downgrade-proof protocol extension to mitigate the underlying vulnerabilities.

\end{abstract}

\maketitle

\section{Introduction}

Electric vehicles (EVs) are rapidly being adopted by consumers and critical transportation services (ambulance~\cite{ev_fleet_nhs}, logistics~\cite{ev_fleet_ups}, public transport~\cite{ev_bus}) alike.
As part of this adoption, DC fast charging has become essential for users who require high availability and fast refueling.
In order to benefit from an increasing network effect and simplify the user experience, industry and governments have jointly pushed for standardization of EV charging technologies.

The Combined Charging System (CCS)~\cite{std_CCS} is the most widely used EV DC charging system, with strong regulatory support from the European Parliament~\cite{ccs_law} and the US Government~\cite{us_nevi}.
The ISO 15118~\cite{std_ISO_15118_2} standards, which define CCS are also used with a different physical connector by the North American Charging Standard (NACS), the successor of the Tesla Supercharger~\cite{std_NACS}.
Additionally, ISO 15118 also supports AC charging, and all newly installed or renovated publicly accessible chargers in the EU will be required to support it from January 2026~\cite{15118_law}.
Therefore, the security properties of ISO 15118 are essential to all modern EV charging systems.

As part of CCS, the EV and the Supply Equipment (EVSE) communicate using a complex protocol stack built on top of a versatile IP/Ethernet link.
This traffic is carried via the widely used Power-Line Communication (PLC) technology called HomePlug Green PHY (HPGP)~\cite{std_HPGP}. 
The use of a well established Ethernet link for communication provides capacity for additional services, such as automatic billing and demand-response charging, in addition to essential communications, i.e., negotiation of charging current and voltage~\cite{std_ISO_15118_1}. 

The security of CCS is intrinsically linked to the security of the communication link, since it carries safety-critical information about the state of the battery, and may also carry personal information.
Researchers have suggested potential avenues for attacks~\cite{conti2022evexchange, low2025security}, but no experiments have demonstrated tampering with a CCS link in a realistic, wireless setting.
Despite using a wired channel, it is known that PLC, as used in CCS charging, unintentionally leaks communication signals via the charging cable~\cite{baker2019losing} and it has been shown in other settings that it is also susceptible to wireless interference~\cite{lampe2016power, kohler2023brokenwire}.

Motivated by the global use of CCS-based charging communications, and the existence of this unintended wireless channel, in this paper we study how an attacker might be able to wirelessly access the communications within the cable, and what impact they can achieve.
Commercial HPGP modems are artificially limited to stop adversarial sniffing and injection of data, and their design is optimized for a wired use-case, limiting their wireless range.

We overcome these limitations by making the following contributions:
\begin{itemize}
    \item Develop a Software-Defined Radio (SDR) implementation of HPGP, capable of wirelessly intruding into CCS communications and granting unrestricted, real-time, and bi-directional access to the physical layer.
    \item Define the practical timing requirements and limitations for a successful hijacking attack, by analyzing 2,750 real-world charging sessions.
    \item Introduce and evaluate new techniques to loosen timing requirements, and greatly improve session hijacking reliability.
    \item Demonstrate fine-grained control over the security level and protocols used by CCS, showcasing that the attacker can easily bypass the use of Transport Layer Security (TLS).
    \item Present and analyze the first wireless end-to-end Man-in-the-Middle attack against the safety-critical power-delivery messages between an EV and charger.
    \item Propose a novel, backwards-compatible yet downgrade-re\-sis\-tant countermeasure which can easily be integrated into existing systems, potentially by just a software update. 
\end{itemize}

\section{CCS Security Overview}

\begin{figure}
    \centering
    \includegraphics[width=\linewidth,trim={0cm 6.8cm 0cm 0cm},clip]{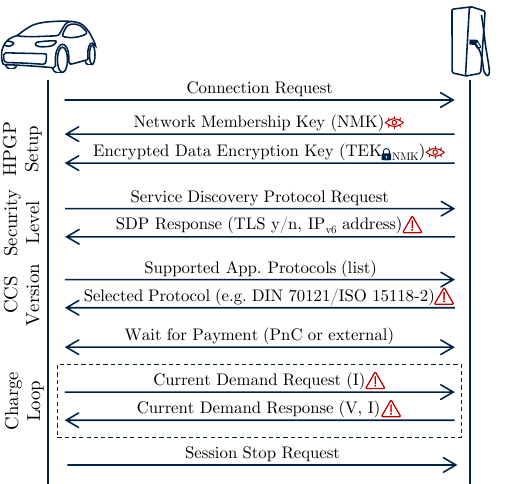}
    \caption{Simplified view of the CCS protocol, highlighting important values that an attacker may extract or control.}
    \Description{A protocol diagram between EV and charger. The stages are HPGP Setup, Security Level, CCS Version, Charge Loop. The risks are leaking the NMK and TEK, editing SDP, editing supported protocol, editing current demand.}
    \label{fig:ccs_protocol}
\end{figure}

In this section, we provide detailed background about the communication stack used by CCS.
A simplified protocol diagram of a CCS charging session is shown in Figure~\ref{fig:ccs_protocol}, highlighting the important stages of the protocol as well as security weaknesses, which are described in detail below.

\subsection{HomePlug Physical Layer}

There are many systematic similarities between HomePlug and Wi-Fi, since they both aim to transmit Ethernet frames in a shared, noisy environment.
The original use case for HomePlug was to create an Ethernet network in a household setting, by sending data between modems over a power line.
Instead of supporting just point-to-point links, HPGP modems form a network in which they can all communicate.
From a network topology point of view, the power lines act not only as the physical medium, but also as an Ethernet switch.

Much like modern RF protocols, HPGP transmits discrete packets using OFDM modulation with robust forward error correction.
Since it uses the very low frequency range of \SIrange{2}{28}{\mega \hertz}, the signal can be transmitted using any wires, without needing specialized RF transmission lines.

To create secure networks over a shared medium, a pre-shared Network Membership Key (NMK) is used.
Devices automatically pair into networks based on this shared key, and exchange ephemer\-al Temporary Encryption Keys (TEKs).
The TEK is sent encrypted by the NMK, which serves both to protect it from eavesdroppers, as well as to authenticate knowledge of the shared NMK.
All user data is then encrypted by the TEK, protecting it from eavesdroppers who do not have access to the keys.
As the NMK is the basis for all encryption, the standard contains multiple key exchange protocols.
Among these, the EV charging ecosystem exclusively uses the Signal Level Attenuation Characterization (SLAC) process described in the following.

\subsection{The SLAC Process}

Because of the potential for HPGP signals to unintentionally couple onto nearby wires due to crosstalk, a vehicle could inadvertently communicate with the wrong charger or even multiple chargers at the same time~\cite{harper2013development}.
To overcome this, the SLAC process measures the signal attenuation between the vehicle and the chargers and selects the charger with the strongest signal.
The selected charger then sends its NMK in cleartext to the vehicle, so that they can begin the usual HPGP pairing process.
Because the SLAC process takes place before the NMK is shared, all packets are transmitted unencrypted. 
This enables a passive eavesdropper with a sufficiently promiscuous receiver to capture the keys, and decode all HPGP traffic~\cite{baker2019losing}.

\subsection{Service Discovery Protocol}

After the EV and EVSE form an HPGP network, the rest of CCS communications use conventional IPv6 Transmission Control Protocol (TCP) and User Datagram Protocol (UDP) traffic.
Both devices assign themselves a link-local IPv6 address, which usually follows the EUI64 format.
To discover the IP addresses, the EV begins by sending a multicast UDP Service Discovery Protocol (SDP) request.
The charger responds to this, indicating the IP address and port of the TCP server (which may be different from the response source address).
The EV connects to this TCP socket, which is used for all further communication.

The SDP process is essential to the security of CCS, because it can be exploited in the same way as ARP spoofing, i.e., to redirect the EV to an attacker controlled IP address.
Crucially, the EV indicates in its SDP request whether it would like to use Transport Layer Security (TLS) or a cleartext TCP connection for the remainder of the protocol.
The EVSE takes this into consideration, and informs the EV what type of socket it is able to provide.
As these messages use standard UDP sockets with no cryptography, an attacker can tamper with the SDP process without being detected.

\subsection{Transport Layer Security}

The earliest version of CCS standardized as DIN SPEC 70121, does not use TLS.
The newer ISO 15118-2 introduces optional one-way TLS, where the EVSE acts as the server.
Finally, ISO 15118-20 introduced mandatory mutual TLS, where the EV and EVSE both authenticate each other.
However, these are not currently widely used, and all deployed devices continue to support non-TLS sessions~\cite{szakaly2025current}.

The Plug and Charge (PnC) protocol allows the EV to automatically pay for the charging session, using certificates.
Due to the sensitive nature of this process, it is restricted to only TLS connections, and the public key infrastructure is fundamentally tied to the PnC payment networks.
There is no list of globally trusted Certificate Authorities (CAs), and CAs are only installed into vehicles when they are added into a PnC network.
Because of this, vehicles with the necessary software for TLS must use unencrypted sessions until they are enrolled into a network, and even then they will only be able to use TLS within their own PnC provider.
Finally, a recent study found that there is no way for a vehicle to tie a specific TLS certificate to a specific charger, allowing any valid certificate from within the same CA network to be used globally~\cite{szakaly2025current}.

\subsection{Vehicle-to-Grid Protocol}

After the TCP or TLS connection is established, the devices communicate using the Vehicle-to-Grid (V2G) protocol.
Over the years, multiple versions have been defined: DIN 70121 (``DIN''), ISO 15118-2 (``-2'') and ISO 15118-20 (``-20'').
To negotiate between them, the EV sends a list of supported protocols and the charger selects one of them based on its capabilities.

After this negotiation, messages are exchanged using the protocol-specific format, but the general flow is the same between all current CCS versions.
As part of this handshake, the EV is given a list of available services to choose from.
Although in most cases the only option is to charge, Value Added Services (VAS) can also be offered as extra features.
VAS currently defined include ``Internet'', where the vehicle can connect to the wider world via the charging cable, or ``Parking''~\cite{std_ISO_15118_20}.
Finally, both sides negotiate payment, either waiting for external user action at the EVSE or by using PnC.

The final and most important part of the V2G communication is the charging phase.
During this phase, the EV and charger communicate to determine the maximum voltage, current, and power each side can support.
Specifically, the EV requests a specific current from the charger, which adjusts its output to meet the request.
The charger acknowledges there requests by also reporting the voltage and current it is supplying.
These current requests are repeated periodically to ensure that the communication link is active, and to adapt to changing circumstances.
As the battery charging process nears the end, the EV slowly ramps down its current request and eventually terminates the charging session.

\section{Threat Model}

Our work focuses on how an attacker can perform a wireless attack against the HPGP charging communications, and use this to interfere with essential parts of the CCS protocol.

\subsection{Capabilities}

We assume an attacker with physical proximity to the charger, but without direct access to tamper with any of the hardware.
Based on this proximity requirement, we consider only attacks that can be performed via wireless communications.
The attacker has access to COTS hardware, such as personal computers, RF equipment or PLC modems, the relevant standards documents, and open-source software.
Finally, we assume an attacker with an SDR implementation of HPGP, giving them full physical layer access, though we later contrast what an attacker could do with a simpler off-the-shelf modem.
We prove the viability of this threat model by presenting our own SDR implementation, created using the public HPGP standard~\cite{std_HPGP}.

\subsection{Goals}

The attacker can have a variety of goals using their MitM position, targeting both the CCS protocol and the entire network interface.
These can include denial of service, sabotage, theft of personal information, or financial gain.
In particular, an attacker may be interested in the messages relating to the power delivery process.
In the best case, such attacks would disable charging as the devices detect a safety concern, while in the worst case they could catastrophically damage the EV battery pack.
In this work we experimentally evaluate such EV charging attacks, and outline other possible attack targets in the Discussion section.

\section{HomePlug SDR System}

\begin{figure*}[t]
    \centering
    \includegraphics[width=\textwidth,trim={0cm 14.5cm 8.7cm 0cm},clip]{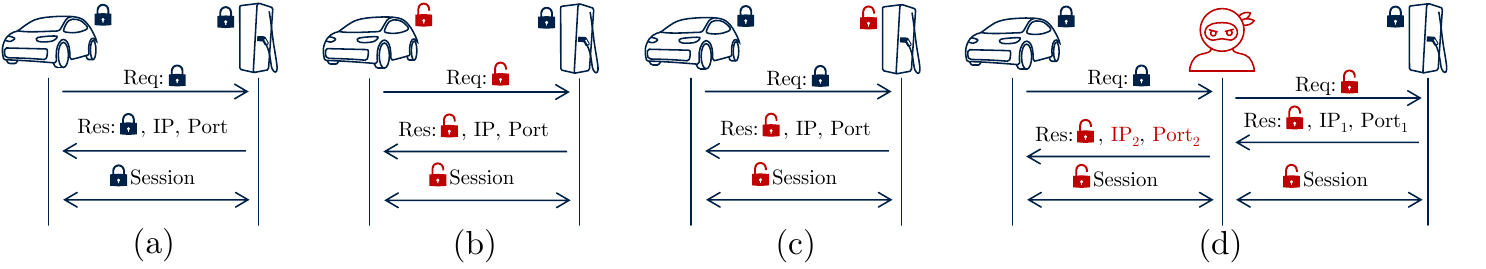}
    \caption{Security negotiation in the SDP process. (a) Two secure devices connecting. (b) A vehicle requesting an unencrypted session. (c) A charger causing a downgrade. (d) An MitM attacker exploiting legitimate behavior to downgrade the connection between two TLS capable devices. To both parties, the observed behavior is identical to the benign cases in (b) and (c).}
    \Description{}
    \label{fig:sdp_hijack_basic}
\end{figure*}

First, we briefly describe our SDR system, the first full-duplex SDR implementation of HomePlug Green PHY.
The system fully implements the modulation and demodulation pipeline, converting seamlessly between analogue waveforms and Ethernet frames.
It also exposes internal stages of the pipeline, allowing us to see and create packets which would not be possible using a commercial modem.
The SDR system does not act as a standalone, albeit software-defined, HPGP modem.
Instead, it spoofs and injects frames into existing connections, allowing a much stealthier operation.

We decided to create an SDR implementation for this work for two reasons.
First, commercial HPGP modems have restrictive firmware, that limits the packets they can send or receive, to those allowed by the protocol.
However, many attacks will rely on the ability to sniff packets intended for different destinations, or forging otherwise impossible packets.

Second, commercial HPGP modems only expose a combined TX/RX connection.
This is convenient for the intended wired use case, where this can be directly connected to the power lines.
However, adding additional amplifiers is necessary to extend the range, and is therefore essential for wireless attacks.
For modems with the combined TX/RX port, this is not possible, since the RX amplifier would directly feed the TX power amplifier, causing a loop.
A duplexer allows a joint TX/RX port to be separated without such loopback, however they are challenging to make for low frequencies and large fractional bandwidths.
We contacted multiple RF device manufacturers, who stated that this cannot be made for the \SIrange{2}{28}{\mega \hertz} frequency range used by HomePlug.
Therefore, this greatly limits the wireless range of a commercial modem.

\subsection{RF Setup}

Our SDR processing is based on a \SI{75}{MSps} data stream, as specified in the standard, and all further processing is done purely in software.
Due to the unique frequency range, we used a USRP N210 with the Basic RX and Basic TX daughterboards for this analogue interface.
The USRP has separate TX and RX connectors, allowing us to connect amplifiers.
The TX port is connected to a \SI{6}{\watt} power amplifier, and uses a dipole antenna approximately \SI{4}{\meter} long, which primarily targets the upper half of the HPGP frequency range.
For reception, we use an off-the-shelf \SIrange{0.5}{30}{\mega \hertz} loop antenna, with an integrated low noise amplifier.

During our lab evaluation, using a commercial portable DC charger, and an EV emulator, we could achieve reliable bidirectional communication at a range of \SI{1.5}{\meter} from the charging cable.
Technical improvements to the antennas and amplifiers could undoubtedly extend this range.

\subsection{Performance} 
Our SDR implementation runs on any modern consumer PC or laptop in real time.
Our testing was conducted on a Kubuntu 24.04 LTS laptop with an Intel i7-1370P CPU, of which the SDR software consumes about 10\%, as reported by \texttt{top}.

The total roundtrip processing delay for the SDR is approximately \SI{12.6}{\milli \second}, measured as the time between the start of a request packet and corresponding response on the wire.
Specifically, the tests were made with SDP packets, measuring the time between an SDP request, and corresponding response.
Therefore, this includes the SDR sampling, demodulation, UDP stack, modulation, and sample output delays.
This can undoubtedly be optimized further, however this delay is well within the timeouts used by CCS, ensuring that our system can reliably participate in the V2G protocol.
We study the specific timing requirements of CCS attacks later in this work.

\section{Active Attacks against CCS}

Sophisticated EV charging attacks rely on the ability to consistently alter all communication between the EV and charger, i.e., an attacker with full MitM control.
However, simply joining the HomePlug network does not grant the necessary MitM access, since packets can pass between the EV and charger unmodified.
Without the ability to cut the charging cable and prevent this direct path, our attack must instead exploit higher layers of the protocol to cause the two legitimate devices to ignore each other.
This problem is shared by other wireless systems such as Wi-Fi, where attackers use ARP spoofing to reroute traffic to themselves and gain full MitM control~\cite{fleck2001wireless}.
ARP is not used in CCS, instead SDP and IPv6 Neighbor Discovery processes are used in its place.
Due to the unauthenticated nature of these protocols, a maliciously crafted packet injected at the right time can be used to spoof the outcome of the process and direct subsequent communications to an attacker controlled IP/MAC address.

By executing this attack, we create two separate TCP streams, one vehicle$\leftrightarrow$attacker, and one attacker$\leftrightarrow$charger.
Data can now be forwarded with arbitrary modifications between the two devices.
In the following, we describe and evaluate in detail how to perform such an attack and study the impact of modifying the power delivery messages.

\subsection{SDP Attack Theory}

As shown in Figure~\ref{fig:sdp_hijack_basic}, the design of the SDP process is fundamentally vulnerable to a hijacking attack.
In the protocol, the vehicle sends a multicast SDP request indicating whether it supports TLS to the charger, which responds whether they jointly support it.
The exchange between two TLS supporting devices is shown in Figure~\ref{fig:sdp_hijack_basic}a.
Unfortunately, there are currently many vehicles which do not support TLS, and all chargers will offer an unencrypted session if asked for one (Figure~\ref{fig:sdp_hijack_basic}b) \cite{szakaly2025current}.
The vehicle will accept this response, as it is also common for chargers without TLS support to downgrade the security level (illustrated in Figure~\ref{fig:sdp_hijack_basic}c).

By combining these legitimate downgrade scenarios, an attacker can replace the vehicle's request to ask for no security (Figure~\ref{fig:sdp_hijack_basic}d), which the charger will accept and follow.
As a result, even though both devices would prefer TLS, the vehicle will believe that it is connected to an insecure charger, the charger will believe that it is charging an insecure vehicle, and the connection will be unencrypted~\cite{zhdanova2022local}.

More importantly, the attacker can send their own IP address in the SDP response to the vehicle, which will thus connect directly to the attacker.
Likewise, the attacker can initiate a TCP connection to the real charger, and relay the traffic (with appropriate modifications) between the two.
The IP stacks within the two devices will cause them to ignore the traffic corresponding to the other connection, and therefore the attacker has achieved an MitM position for the remainder of the V2G communication session.
However, this attack against SDP once again assumes that the attacker is able to modify the messages, which is not possible.
Instead, they must rely on speed to inject a new packet into the communications between the EV and charger at the right time.
We therefore need to evaluate the SDP implementations within real devices, to understand how exactly this attack can be done, and what timings are required.

\subsection{Charger Response Time Analysis}

\begin{figure}[t]
    \centering
    \includegraphics[width=0.9\linewidth,trim={0cm 7.2cm 0cm 0cm},clip]{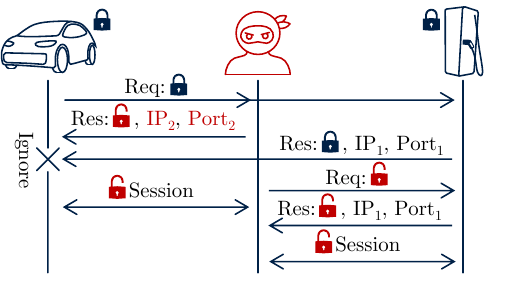}
    \caption{SDP hijacking by racing the responses. The attacker can augment their chances by sending various attacks to the charger before the legitimate SDP response is sent.}
    \Description{}
    \label{fig:sdp_hijack_race}
\end{figure}

The SDP implementation within all EVs we observed is straightforward: they periodically send multicast SDP requests until the first response is received, and then connect to the address indicated in the response.
Therefore, the attacker must send a reply with their own IP address and preferred security level before the charger does so.
We illustrate the process for a timing based SDP attack in Figure~\ref{fig:sdp_hijack_race}.
This race condition is a challenge for the attacker, who must beat the speed of highly optimized HPGP chip and charger implementations.
Therefore, we later discuss several necessary techniques to gain a considerable advantage on the legitimate system.

To better understand the performance of real-world chargers first, we analyze a large dataset of real-world charging sessions from prior work~\cite{szakaly2025current} to extract exact information on the speed of an SDP response.
The dataset contains packet captures taken by an EV emulator connected to real chargers, and we extract the time between the SDP request being transmitted and responses received by the vehicle.
The captures were taken at the OS level using \texttt{tcpdump}, hence the measured time excludes the EV emulator's processing time.
The measured delay includes the EV-side Linux kernel, EV HPGP modem, charger HPGP modem, and charger software stack.

The entire dataset at the time of analysis contained 2,750 successful SDP exchanges.
Within the data, there are clear differences between brands, likely due to different SDP software implementations.
For the purposes of this analysis, we consider the dataset to contain a good cross section of different chargers, and thus the statistics observed are representative of a typical real-world charger.

\begin{figure}
    \centering
    \includegraphics[width=\linewidth]{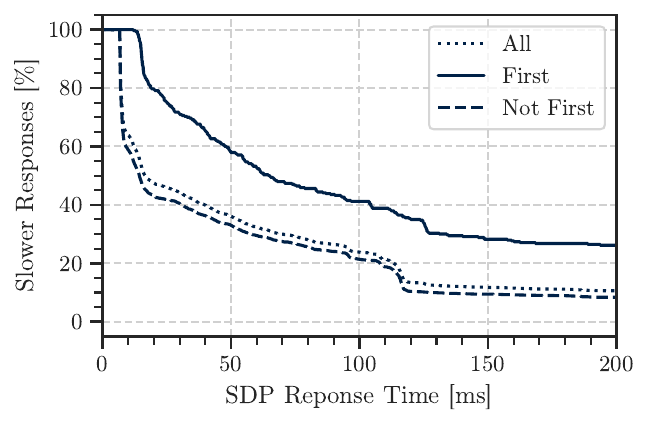}
    \caption{Speed of SDP responses in real charging sessions, extracted from the dataset in~\cite{szakaly2025current}. The results show that First exchanges are consistently slower.}
    \Description{}
    \label{fig:sdp_times}
\end{figure}

More importantly, we observed a systematic difference between the first SDP exchange performed with the charger and subsequent exchanges.
Specifically, first exchanges were consistently slower than later ones on the same charger.
The tail distribution function of SDP roundtrip times based on the dataset is shown in Figure~\ref{fig:sdp_times}.

We found that the fastest SDP roundtrip in the dataset was \SI{6.2}{\milli \second}, while the fastest first one was \SI{11.9}{\milli \second}.
Additionally, \SI{95}{\percent} of the first exchanges were slower than \SI{14.9}{\milli \second}, while only \SI{49}{\percent} of the subsequent exchanges were slower than this.

Deeper analysis of the packet captures reveals that this behavior is caused by the IPv6 stack inside the charger, which must determine the MAC address of the EV before sending the SDP response packet.
While this information is contained in the source MAC field of the SDP request, compliant IPv6 implementations are forbidden from inferring it this way~\cite{rfc4861}.
Instead, it must only be obtained via the ICMPv6 Neighbor Discovery process, which adds an additional roundtrip before the SDP response can be sent, and explains the almost exact doubling in the fastest measured responses.
Subsequent SDP exchanges with the same charger-vehicle pair will be faster, as the information is stored in the charger's Neighbor Discovery Cache for up to a few hours.

Using the same methodology as this dataset, we measured the average SDP response time of our SDR to be \SI{16.1}{\milli \second}, which is sufficient to attack \SI{85}{\percent} of real-world chargers on the first attempt, posing a serious risk to much of the EV charging infrastructure.
Additional optimization of the SDR implementation could undoubtedly improve this response speed, and improve the success rate of this timing-based attack.
However, instead of creating such a highly fine-tuned SDR, we believe that there is more scientific value in understanding what an attacker can do with a reasonable implementation.
Therefore, we now study advanced attack techniques, which greatly extend, or even eliminate the time window, ensuring constant and reliable attack success.

\subsection{Advanced SDP Attack Techniques}

To improve attack reliability, we take advantage of the necessary implementation details of SDP and the underlying IPv6 standard.
First, we consider what chargers do when receiving multiple different SDP requests, i.e. the legitimate request from the vehicle, as well as one from the attacker.
We found that in some cases, chargers stop responding to SDP requests once the TCP connection was set up, and in some cases they only respond to the first request.

\subsubsection{Pre-Hijacking}
\begin{figure}[t]
    \centering
    \includegraphics[width=0.9\linewidth,trim={0cm 7.2cm 0cm 0cm},clip]{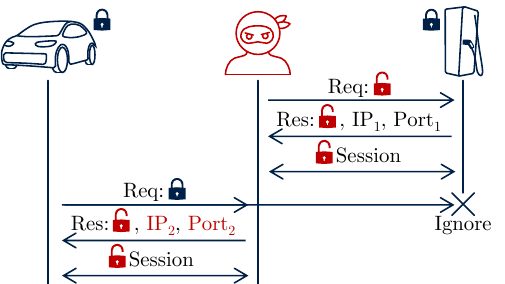}
    \caption{The SDP pre-hijacking attack, applicable against chargers which ignore SDP requests after the connection has been established.}
    \Description{}
    \label{fig:sdp_hijack_pre}
\end{figure}

In such cases, to successfully downgrade the security level the attacker must perform an SDP exchange with the charger \textit{before} the vehicle does.
As a benefit to the attacker, the charger will entirely ignore and not respond to any subsequent SDP requests coming from the EV, eliminating the need to race responses.
We illustrate this attack in Figure~\ref{fig:sdp_hijack_pre}.

To execute this pre-hijacking technique, our SDR implementation has a major timing advantage over the legitimate EV, since it does not need to formally join the HPGP network to start communicating.
Instead, immediately after the encryption keys have been transmitted, our implementation can begin injecting and sniffing packets, and thereby communicate with the charger.
In contrast, the EV's modem needs to exchange additional management frames to formally join the network, before it can signal to the vehicles software that it is ready to communicate.

To precisely understand the performance of real-world HPGP modems that we want to pre-hijack, we measure the time between the TEK (data encryption key) being sent on the wire, and packets being sent by the various devices.
We collected 131 exchanges from commercial devices, including ones with and without the MitM active.
This included two different EV models from entirely unrelated manufacturers %
and a commercial EV emulator,%
as well as chargers from three different manufacturers. %

\label{sec:join_speed}
Detailed analysis and a plot of this data is presented in Appendix~\ref{app:join_speed}.
In summary, the first SDP request from any EV side modem was sent \SI{344.3}{\milli \second} after the TEK, while our MitM could do the same in \SI{12.3}{\milli \second}.
This large difference shows that our SDR has a considerable timing advantage in communicating with the charger, well before the EV begins the SDP process.
We assume that this difference is because the EV-side HPGP modem needs to exchange multiple management messages, before informing the EV charge controller that it is ready.
This means that any attack that requires ``preparing'' the charger, including the pre-hijacking attack, is guaranteed to happen in time using our SDR implementation.

\subsubsection{Neighbor Discovery Attacks}

Motivated by the possibility to attack the charger before the EV starts SDP, we now consider attacks to improve the success rate of the SDP response race condition.
Recall that our charger timing analysis revealed that chargers require the IPv6 Neighbour Discovery (ND) process to complete before sending the SDP response.
In addition to giving the attacker additional time to inject a spoofed response packet, the ND process is itself vulnerable to various spoofing attacks.
Specifically, an attacker can use Neighbor Advertisement (NA) spoofing to modify the IP\textrightarrow MAC address mapping, and cause packets to be sent to an attacker supplied MAC address.
The HPGP layer will still route these packets to the vehicle, but compliant Ethernet implementations will discard them based on the destination MAC address field.
Prior work has suggested NA spoofing as a potential technique, but did not evaluate it~\cite{eder2025charging}.
In our analysis, we found two key challenges with this method, and present ways to overcome them.

First, sending NA packets before the SDP process begins does not have an effect.
This is because NA packets cannot create new entries in the neighbor cache, they can only change the current values (following RFC 4861~\cite{rfc4861}).
Entries are only added to the cache when the victim OS wants to send a packet to a device, and this only happens after it has received the SDP request.
It is theoretically feasible to send an NA packet in the narrow time window between the request and the response, but if this could be done reliably, one could simply send the SDP response instead.

Instead, our novel technique is to send an ICMPv6 Echo Request (ping) to the charger, immediately followed by the spoofed NA packet.
This process is illustrated in Figure~\ref{fig:na_attack} in the Appendix.
While processing the ping and generating a response, the victim inadvertently creates the cache entry, which can then be overwritten by our NA.
Notably, we can anticipate that the device will send an ND request upon receiving the ping packet, without needing to wait for this roundtrip.
The ISO 15118-2 standard requires that all devices implement IPv6 Ping and NA packets, therefore this technique is expected to work in all compliant chargers.

Second, we discovered that several EVs use a simplified IP stack and that they do not validate the destination MAC address of received packets.
Therefore, even though NA spoofing successfully caused the SDP response to be sent to a non-existent MAC address, the EV still received and processed it normally.
To overcome this challenge, we set our NA packets to advertise that the IP address of the vehicle is located at the MAC address of the charger.
When the charger generates the response packet, it is thus addressed to its own MAC address, and therefore the packet is dropped before it ever reaches the wire.
We believe that a device accepting NA packets for its own MAC address is not a bug within its IP stack, but instead a necessary part of the IPv6 Neighbor Proxy mechanism.
Therefore, we believe that this attack does not target implementation flaws, but issues within the standards.

\subsection{MitM Connection Establishment}

\begin{table}
    \resizebox{.5\textwidth}{!}{
    \begin{tabular}{lccccc}\toprule
    Device & Power Rating & Type & DIN & 15118-2 & TLS\\\midrule
    \fullX & \SI{475}{\volt}, \SI{375}{\ampere}& AC\&DC & \ding{51} & \ding{51} & \ding{55}\\
    \fullY & \SI{412}{\volt}, \SI{165}{\ampere} & AC\&DC & \ding{51} & \ding{55} & \ding{55}\\
    EV Emulator~\cite{szakaly2025current} & -- & -- & \ding{51} & \ding{51} & \ding{51}\\
    \midrule
    \fullA & \SI{500}{\volt}, \SI{20}{\ampere} & DC & \ding{51} & \ding{55} & \ding{55}\\ %
    \fullB & \SI{500}{\volt}, \SI{375}{\ampere} & DC & \ding{51} & \ding{51} & \ding{55}\\ %
    \fullC & -- & AC & \ding{51} & \ding{51} & \ding{51}\\
    \fullD & \SI{453}{\volt}, \SI{125}{\ampere} & DC & \ding{51} & \ding{55} & \ding{55}\\
    \bottomrule
    \end{tabular}
    }
    \caption{Summary of the different device types used during our evaluation. The devices are all from unrelated manufacturers. Different instances identical devices were used.}
    \label{tab:devs}
\end{table}

We tested the above attacks on four unrelated EV charger brands, including chargers from two of the largest manufacturers in both Europe and the US.
The list of all devices used during these experiments is summarized in Table~\ref{tab:devs}.
In all cases, we found that IPv6 NA Spoofing or Pre-Hijacking were able to achieve a reliable attack.

\subsection{MitM Attacks on TLS}

Although TLS is currently not widely used by EV chargers~\cite{szakaly2025current}, we expect that all EVs and chargers will support it at some point in the future.
However, it is likely that insecure sessions will continue to be supported for a long time, opening the possibility for TLS stripping attacks.
Therefore, we experimentally confirm that the TLS stripping attack that we proposed above works on real hardware.

\paragraph{Setup}
For this testing, we used \fullC, a PnC enabled, real commercial charger which necessarily supported TLS but also accepted unencrypted sessions.
The charger was set up by the manufacturer, and configured with their certificates.
As a commercial device, we had no ability to reconfigure it from these settings.
We also used an open-source vehicle testing emulator~\cite{szakaly2025current}, which was configured to request a TLS session, but allowed downgrades.
Both of these configurations are in-line with modern real-world devices, which must maintain compatibility with devices which do not support TLS.

\paragraph{Evaluation}
Using passive eavesdropping, we verified that under normal circumstances the two devices connect using TLS, which protected the remaining messages.
Next, using our MitM setup, we performed the above SDP hijacking, aided by NA spoofing, and injected SDP packets indicating no TLS should be used.
Both sides accepted this downgrade, and we opened unencrypted sockets towards the car and charger.
As all encryption was stripped, the MitM had access to the communications that were previously encrypted.

Using this same setup, we also demonstrated an asymmetric TLS downgrade, where we requested a TLS connection from the charger, but downgraded the vehicle.
Since in ISO 15118-2 the charger acts as the TLS server, this attack can be performed by the attacker without requiring any certificate.
The result of this attack was a TLS connection between the charger and MitM, but no encryption between EV and MitM.
Since the protocol within a TLS connection is identical to the unencrypted one, we could then forward the data as raw binary.

\paragraph{Discussion}
Such an asymmetric attack may be useful in future cases where chargers requires TLS, but vehicles still accept insecure connections.
Additionally, an attacker may want to target features within the charger that are only exposed during a TLS session, such as values added services, without needing to obtain a certificate trusted by the vehicle.
Finally, even once vehicles require TLS, if the attacker can obtain a single trusted certificate, they can use this to MitM all vehicles globally who trust the root certificate~\cite{szakaly2025current}.

\subsection{Protocol Version Attacks}

The first messages within the V2G communication channel negotiate the preferred version of the protocol (DIN, 15118-2 or -20).
With MitM access, the attacker can change the list of requested versions to offer only a single protocol to the charger, and thereby force the version used.

\paragraph{Setup}
To demonstrate this capability, we tested on \fullX and \fullC.
We had no administrative access to either of these devices, so they were using their factory settings.
Both devices supported DIN and ISO 15118-2, but the vehicle marked DIN as the preferred version during the negotiation.

\paragraph{Evaluation}
First, by passive sniffing, we observed that the two devices negotiated to not use TLS because of the vehicle, and then chose the DIN protocol as this was preferred.
Next, we performed an MitM attack aided by NA spoofing, and at the SDP stage we selected a non-TLS connection.
During protocol negotiation, we replaced the request coming from the vehicle to only indicate support for ISO 15118-2.
With no other option, the charger accepted this request.
We confirmed to the vehicle that ISO 15118-2 was selected, after which we forwarded the messages between the two as raw binary.
Therefore, we successfully took two devices which natively communicated using DIN, and spoofed version negotiation to select ISO 15118-2 instead.

\paragraph{Discussion}
The benefits of such an attack could be to exploit flaws present in only one version of the protocol.
This could be an implementation bug in a specific version, or a vulnerable feature introduced by a new version.
Our finding that the vehicle prefers DIN highlight that such a version attack does not strictly have to downgrade, but it can also be used to upgrade the protocol version.
Additionally, much like with TLS, the attacker can use different protocol versions with the two devices.
In this case, the attacker cannot directly forward binary messages, but they must re-encode each packet in the appropriate format.
Since the various CCS versions to date largely follow the same protocol flow and message content, this could be implemented without difficulty.

\section{Power Delivery Attacks}

The fundamental risk posed by an attack against CCS charging communications is that the attacker can tamper with the safety-critical power delivery messages.
In regularly sent \currentdemandreq messages, the EV requests a specific current from the charger, based on the actual battery conditions.
This changes during charging, decreasing to 0 as the battery becomes fully charged.
Editing these messages during a charging session could degrade performance, or pose a risk to the safety of EV charging as a whole.

To each request message, the charger responds with a \currentdemandres, which indicates that the connection is still active.
Additionally, the response contains the voltage and current present on the wire, as measured by the charger.
This poses a potential risk for confusion within the EV charge controller, since the EV could make its own measurements of these values, or trust the numbers provided by the charger.
Principally, any secure implementation should compare both values, and terminate the charging session in case of any discrepancy.
However, prior work has shown that it is possible to draw power from the battery of many EVs, while telling it that it is being charged~\cite{low2025draindead}.
Allowing a negative current to flow through the charging port on a vehicle that does not support bi-directional charging indicates a serious lack of validation by the charge controller, that may be exploitable in worse ways.

\begin{table}
    \begin{tabular}{cccc}\toprule
    Vehicle&Charger&Real&Spoofed\\\midrule
    \shortX&\shortB&\SI{440}{\volt}, \SI{20}{\ampere}&\SI{0}{\volt}, \SI{0}{\ampere}\\
    &\shortA&\SI{448}{\volt}, \SI{86}{\ampere}&\SI{1330}{\volt}, \SI{1330}{\ampere}\\
    \shortY&\shortD&\SI{371}{\volt}, \SI{112}{\ampere}&\SI{240}{\volt}, \SI{40}{\ampere}\\
    &\shortD&\SI{373}{\volt}, \SI{105}{\ampere}&\SI{670}{\volt}, \SI{570}{\ampere}\\
    \bottomrule
    \end{tabular}
    \caption{Tested ranges when spoofing \currentdemandres messages from the charger to the EV.}
    \label{tab:res_spoofing}
\end{table}

To investigate these vulnerabilities and their potential safety impact further, we conducted experiments where we live-edited \texttt{Current\-Demand} messages, using a graphical user interface to show and modify the values for intuitive monitoring.
More detail about our specific implementation is described in Appendix~\ref{app:tool} and Figure~\ref{fig:attack_screen}.
As demonstrated by the above evaluation, we can also control the protocol versions and use of TLS, as we are setting up the MitM channel.

A major risk in power delivery attacks is not only that the attacker can control critical values but that they can continue to communicate with the charger even after the EV requests to terminate charging.
To achieve this, the attacker must implement two independent CCS charging loops, which communicate with the EV and SE independently.
The responses to the EV are generated directly by the attacker and, similarly, the attacker sends requests to the charger.
This also helps to reduce the round-trip time measured by both devices, improving the reliability of a long-term attack.

Initially, the physical values are synchronized, so newly sent messages to the charger use the latest received request from the EV, and vice-versa.
However, the attacker can interactively override the values to perform various experiments.
In our implementation, the attacker can introduce subtle offsets, fixed multipliers, or entirely replace values with a constant.
For safety reasons, we also implemented an emergency stop capability, which sends requests to the charger to immediately terminate power delivery.

\subsection{Power Reporting Attacks}

The EV can use the voltage and current measurements sent by the charger in the \currentdemandres messages to double-check its own measurements, and improve the ongoing safety of the charging session.
However, a worst-case EV implementation could fully trust the charger and rely only on these values, without making any of its own measurements.
To evaluate how EVs process them, we conduct experiments by modifying only these reported values.

\paragraph{Evaluation}
Our evaluation was conducted on two real EVs, using commercial DC chargers with factory settings.
We tested \fullX with \fullA and \shortB; and \fullY with \fullD.
We performed the MitM takeover as described before, selecting DIN 70121 and no TLS for simplicity in all cases.
Initially, we forwarded all values faithfully, however we eventually started modifying the values sent to the EV in \currentdemandres messages.
In all cases, we never modified values sent from the EV to the charger, ensuring that the EV was charged with exactly the requested power.

\begin{figure}
    \centering
    \includegraphics[width=\linewidth]{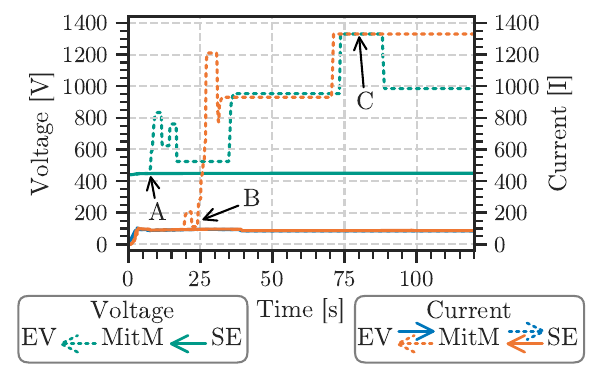}
    \caption{Attack against \fullX, showing edited \currentdemandres messages. The attack begins by increasing the reported voltage (\legendline{cyclergreen}{pythondash}) at {\normalfont \textit{A}}, and current (\legendline{cyclerorange}{pythondash}) at {\normalfont \textit{B}}. Finally, at {\normalfont \textit{C}} the vehicle is told that it is being charged at \SI{1330}{\volt}, \SI{1330}{\ampere}, but in reality it is \SI{448}{\volt}, \SI{87}{\ampere}. The vehicle continues to charge normally despite the attack. This demonstrates real-time, interactive control.}
    \Description{}
    \label{fig:mitm_reporting}
\end{figure}

The lowest and highest spoofed values are summarized in Table~\ref{tab:res_spoofing}.
In short, for both vehicles we sent values which indicated an unreasonably low and high voltage for the battery.
Additionally, we plot the values being sent between the devices during one of the experiment in Figure~\ref{fig:mitm_reporting}.

\paragraph{Results}
Despite the unreasonable reported voltages, which far exceeded the safe operating range for the nominally \SI{400}{\volt} battery packs, neither vehicle tried to stop the charging session, and continued charging normally.
This indicates that the vehicles are relying on their own internal measurements.
However, a large discrepancy between the EV internal measured value and the charger reported value could be caused by a faulty sensor, major losses within the charging cable, or, as in our case, the MitM attack.
In any case, charging cannot safely continue, and therefore we believe that all vehicles should terminate the session immediately.

\begin{figure}
    \centering
    \includegraphics[width=\linewidth,trim={0cm 18cm 0cm 35cm},clip]{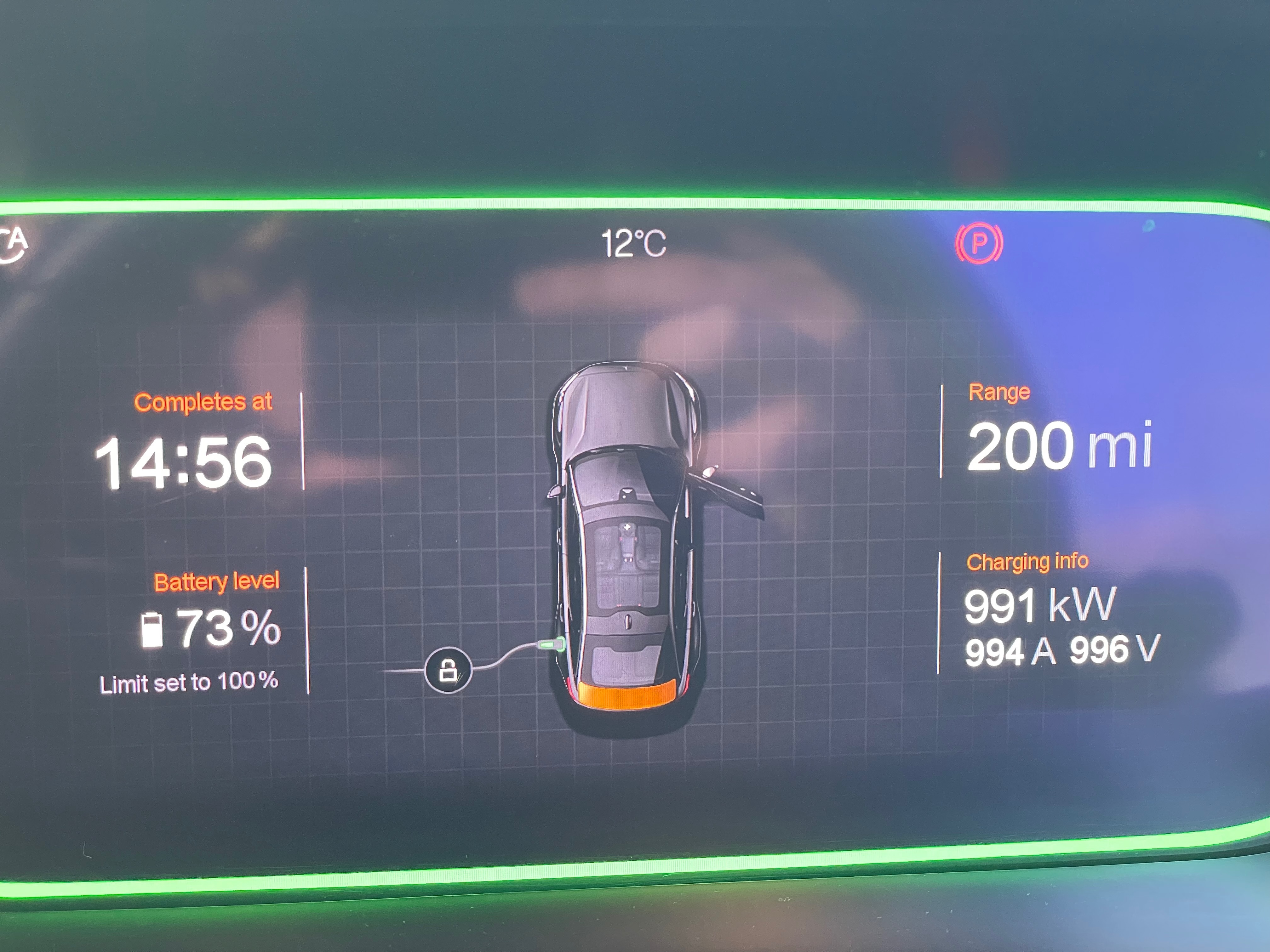} %
    \caption{The dashboard of \fullX during an MitM attack. The vehicle is actually being charged with \SI{91}{\ampere} at \SI{442}{\volt}, however the values set by the attacker in the \texttt{CurrentDemandRes} message are shown on the dashboard.}
    \Description{}
    \label{fig:car_dash}
\end{figure}

While the above behavior would also be consistent with a vehicle that simply ignores the reported values, \fullX  had observable side-effects during the attack.
It displayed the attacker-supplied current and voltage values on the dashboard as shown in Figure~\ref{fig:car_dash}, which means that the externally supplied values are passed further into the vehicles internal network.
Additionally, when the optional \texttt{EVSE\-Power\-Limit} field was set in our spoofed response packet, then \fullX reduced its current request to \texttt{EVSE\-Power\-Limit}/\allowbreak{}\texttt{EVSE\-Present\-Voltage}, where both values were supplied by the charger.
This means that the car adjusts its request such that the requested power does not exceed the chargers power limit.
While this is unnecessary, it clearly shows that the charger supplied values are processed within the charge controller, and actively fed into charging calculations.

\subsection{Power Reduction Attacks}

In the previous experiments, we only modified the informational response messages sent from the charger to the EV, and found that while vehicles use their own sensor measurements, they also process the externally supplied values.
To better understand the risks of charging attacks, we now consider scenarios where the power delivered by the charger is different than requested by the EV.
First, we only lower the power delivered, which should be safe in all cases, as vehicles are designed to accept less power than they requested.
From the charger's perspective, it is simply meeting the demands of an EV which does not support high speed charging, while from the EV's perspective, the charger is under-delivering, which is commonly done by low-power chargers or in cases of limited electricity supply.

There are many potential legitimate reasons behind a slow charging session, such as thermal throttling, power availability on the smart grid, battery degradation, etc.
User interfaces only provide limited information, making it impossible to determine whether the slow charging is caused by the EV's request or the charger's capabilities.
Therefore, the operator will not be able to identify the cause of the slow charging.
Such an attack could be used to effectively deny service, without the obvious signs of a complete denial of service.

\paragraph{Evaluation}
Our evaluation was identical to the previous experiment, using the same vehicles and chargers.
Once again, we performed the attack on DIN 70121 and no TLS.
In this case, after an initial period of normal charging, we lowered the current requested by the vehicle from the charger.

\paragraph{Results}
We found that the attack is successful in all cases.
The charger faithfully follows the request from the vehicle, and decreases the charging power.
The vehicle has no way to detect this as suspicious, and therefore charging continues at a slower speed.
However, we observed an interesting behavioral difference between our two vehicles.

\fullX always requests exactly the current it wants, regardless of what the charger was providing.
Therefore, as we decreased the supplied current, there was no change in the requests.
However, \fullY seems to use a feedback loop, where it increases its request if it wants more current, or reduces it to get less, regardless of the actual values.
This behavior allows it to correct for the case where there is a small systematic error in the charger's output.
Nonetheless, the request remained clamped between \SIrange{0}{125}{\ampere}, which was the maximum charging current for the charger during this experiment.

\subsection{Overcharging Attack}

Conversely, the attacker can also increase the current request.
Batteries have a maximum current limit that they can safely handle without excessive heating, and more importantly, they have a voltage limit above which permanent chemical breakdown of the cells occurs.
Under normal conditions, the EV's charge controller will indicate the safe current to the charger, which will under no circumstances exceed this.
As the voltage reaches the limit, the requested current is slowly reduced to zero, and the EV signals the end of the charging session.

However, an attacker can increase the current requested by the vehicle, and when the EV requests to terminate the charging session, they can continue to communicate with the charger and continue the charging session.
This way, energy keeps flowing into the EV, degrading its lifetime or potentially causing catastrophic damage.

\paragraph{Evaluation}
To test this attack, we used \fullY (a \SI{50}{\kilo \watt} vehicle), and a \fullD (a \SI{50}{\kilo \watt} public DC charger).
The test was conducted with the support of a local and a federal government entity, which supplied the charging system and the vehicle, respectively.
The maximum power of this charger was the same as the rated charging power of the vehicle, limiting the potential danger of these tests.
In order to achieve meaningful overcharging, we waited until the vehicle was nearly full.
At this point, the requested current begins to decrease, and at the time of testing was only \SI{64}{\ampere} out of the maximal \SI{165}{\ampere}.
Once again, the selected protocol for this experiment was DIN 70121 and no TLS.

\begin{figure}
    \centering
    \includegraphics[width=\linewidth]{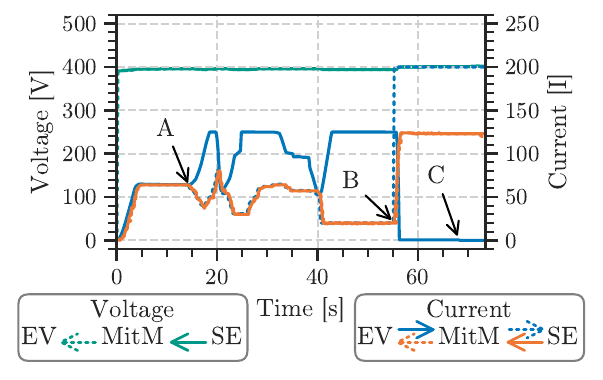}
    \caption{Power delivery attacks against \fullY. At {\normalfont \textit{A}}, we modified the requests to slow down charging (\legendline{cyclerblue}{pythondash}), and the charger reduces its power (\legendline{cyclerorange}{solid}). The vehicle tries to compensate by increasing its request (\legendline{cyclerblue}{solid}). At {\normalfont \textit{B}}, we suddenly ramp the requested power, and begin overcharging.}
    \Description{}
    \label{fig:mitm_overcharge}
\end{figure}

\paragraph{Results}
The charging parameters exchanged during this attack are shown in Figure~\ref{fig:mitm_overcharge}.
We first performed the undercharging attack, and observed the behavior to be as expected.
After some time (point B in the Figure), we modified the MitM to suddenly ramp the request to the maximum, and the charger ramped the output to its maximal value of \SI{124}{\ampere}.
At the same time, the vehicle dropped its request to \SI{0.5}{\ampere}, which was consequently overwritten by the MitM.
This continued for \SI{12}{\second}, after which the vehicle further dropped its request to \SI{0}{\ampere} (point C in the Figure), which again was hidden by the MitM.
The attack continued for an additional \SI{5}{\second}, until the charger terminated the session.

\paragraph{Discussion}

At the end of the attack, the charger continued to receive the same data stream as previously, however, it still chose to terminate the charging session.
This means that the charger received some sort of information from the vehicle that was not blocked by the MitM.
We consider two options:
First, the vehicle can internally pull a contactor to disconnect the battery, which will cause a voltage surge at the charger, as current stops flowing.
Alternatively, the vehicle can use the basic signaling data lines within the charging cable.
Basic signaling does not carry any data stream, just a simple voltage indicating that charging is currently allowed.

In total, the vehicle was charged at double the originally requested current for \SI{17}{\second}.
If the timings of the attack do not depend on the charging power, then a modern high power charger (e.g. \SI{400}{\kilo \watt}) could have delivered a substantial shock to the battery. %
Therefore, starting such an attack near the end of the charging session risks overcharging the battery with unforeseeable consequences.

\subsection{AC Charging Attacks}

ISO 15118-2 extends beyond DC charging, to offer AC charging.
AC charging is a much simpler protocol, as the EVSE (charger) does not perform any voltage conversion.
Instead, the EVSE is simply an electrical outlet, with metering and payment capabilities.
Accordingly, the power delivery messages are much simpler, as the SE informs the vehicle about the available mains voltage (this cannot be changed), and the maximum current that the vehicle can draw without overloading the system.
All AC-DC conversion and charging happens within the vehicle, and therefore the attacker cannot overcharge the battery.

Potential attacks could instead decrease the maximum current available, to perform the covert slow-charging denial of service attack described previously.
Alternatively, the attacker could indicate a higher maximum current than supported by the wiring.
This would overload the electrical supply of the SE, or the charging cable itself.
Such an overload could trip a fuse within the system, or worse.
Additionally, AC charging supports PnC and VAS, and therefore attacks against these systems could equally be launched against ISO 15118-2 AC charging systems.

\paragraph{Evaluation}
To validate that the MitM attacks described in this paper are viable against AC charging systems, we used \fullC, a commercial ISO 15118-2 AC charger, and \fullX.
We confirmed that the MitM can successfully hijack the communication session as before, and gain access to the V2G messages.
For safety reasons, we could not perform the potentially destructive tests against the power delivery portion of the protocol described above, and therefore this is left for future work.

\section{Countermeasures}

In this section, we enumerate potential countermeasures to the attacks presented.
The exploited vulnerabilities are systemic to the entire CCS protocol and found throughout all layers of the stack:
Expanding on cruder vulnerabilities from previous work such as Brokenwire \cite{kohler2023brokenwire} with a more fine-grained technique, the attack breaks through the physical barrier of the wired charging cable using radio waves, defeats the HomePlug encryption due to the publicly broadcast key, and gains access to the trusted IP network between the devices.
It further exploits IPv6 routing using Neighbor Advertisements, skips TLS using the unauthenticated SDP process and targets power delivery messages due to weak sanity checks.

The Combined Charging System is embedded into the hardware, firmware, and software of millions of EVs and chargers around the world.
Therefore, it is not possible to entirely replace the system, and instead we consider improvements that can be introduced as optional for now, and made mandatory after sufficient adoption has been reached.
We consider all layers and also discuss ways in which individual components can protect themselves.

\subsection{Standalone Defenses}

A MitM attack will inevitably create unusual traffic patterns.
Devices should be more vigilant to this, and more carefully validate received data in accordance with the plausible protocol flow.
More specifically, vehicles should look for conflicting SDP responses, and chargers should look for conflicting requests.
This monitoring should continue even after an SDP exchange is considered to be complete.
Multiple messages with different IP addresses or security levels are not possible in normal systems.

EVs should base all charging safety decisions on their own, internal measurements.
They should never rely on values reported by the charger, but they should still be checked for consistency.
Any discrepancy should cause the charging session to terminate due to an error.

Chargers should only ever increase the supplied current slowly, giving the EV time to detect and terminate the charging session in case of any issue.
This limits the attackers ability to ``shock'' the car by suddenly ramping up the current request.
The standard should explicitly mandate and conformance test for such limits.
However, chargers should ramp the current down as fast as safely possible.

\subsection{Alternative Physical Layer}

The HomePlug physical layer is built into all current hardware.
It should be investigated if an alternate PHY system can be introduced. %
The modulation of the new PHY system should not resemble a wireless protocol (e.g. it should not have a carrier frequency), in order to minimize signal leakage issues.

The upcoming Megawatt Charging System (MCS) is a derivative of CCS designed for high-power applications, using the same IPv6 based protocol stack.
After initially considering PLC~\cite{std_MCSv1}, this was replaced by twister pair Ethernet~\cite{std_MCS}.
While MCS itself is not practical for passenger vehicles due to the excessive connector size, the same signaling could be considered for CCS as well.

\subsection{Downgrade Prevention}

The cleartext transmission of the NMK leaves it vulnerable to eavesdropping attackers, and compromises the integrity of the HPGP network.
Cryptography could be used to ensure the confidentiality and integrity of these initial messages.
A secure key-exchange such as Diffie-Hellman would eliminate the risks of a passive eavesdropper, and additional signatures can be used to protect against active attacks, at the cost of introducing a key management problem.
The HPGP standard defines the Secure SLAC process~\cite{std_HPGP}, which introduces public key cryptography to SLAC, however, this is explicitly not used by any existing version of CCS. %

If implemented, Secure SLAC would suffer from a downgrade attack, where the attacker can simply pretend that they do not support it.
Similarly, as long as devices support non-TLS connections, the SDP protocol will allow for the downgrade attack demonstrated in this paper.
Any attempt to introduce optional protocols to improve security will be futile, since the MitM attacker will be able to pretend that they are a legacy device without support, and thereby force a downgrade attack.

The fundamental solution to this issue is to permanently eliminate insecure protocol versions, however, this will break charging for many legacy devices.
Since many vehicles may not have a way to update their charge controller firmware, such a change would need significant government support, and it would have to wait until most old devices have reached the end of their service lives.
Considering that TLS in CCS was introduced by the ISO-15118 standard in 2014, yet in 2024 only a small fraction of devices supported it~\cite{szakaly2025current}, we believe that such a change could take a decade.

\subsubsection{Downgrade-Proof Method}

Motivated by this, we propose a protocol extension that is simultaneously downgrade-proof between devices which support it, and backward compatible with those which do not.
As discussed above, this cannot be done over the potentially attacker controlled channel, so we turn to a separate, existing, but underutilized physical channel between EV and EVSE.

Basic signaling is used by Type-1 and Type-2 AC EV charging, for basic presence detection.
This signal is a \SI{1}{\kilo \hertz}, \SI{\pm12}{\volt} Pulse Width Modulated signal, generated by the charger.
When connected, the vehicle applies a resistor to this signal, lowering the voltage.
This enables a primitive two-way channel, where the charger can change its duty cycle, while the vehicle can change its applied load resistance.
To maintain compatibility, this system still exists within CCS, where the most important duty cycle set by the charger is \SI{5}{\percent}, indicating that HPGP should be used.

Importantly, this system does not have an RF-like packet design, and does not leak wirelessly.
Since we consider that modifying with the charging cable will be visually obvious to users, this channel is safe from the attacker.
Using this, we propose to modify the \SI{5}{\percent} carrier, to contain additional data.
To allow for errors, the standard allows this signal to be in a range of \SIrange{3}{7}{\percent}~\cite{std_IEC_61851_1}.
Therefore, the charger could generate, and the vehicle could detect \SI{4}{\percent} and \SI{6}{\percent} as two separate states, assigned to mean bits 0 and 1.

A charger implementing our scheme would replace the \SI{5}{\percent} carrier at the start of the session with a data transmission, by changing the length (duty cycle) of each pulse sent.
First, alternating 0 and 1 bits would be used as an easily detectable pattern, to signal that the charger supports our extension.
If the vehicle does not detect the initial 01 repeating pattern in the first few cycles of the signal, it can immediately begin the SLAC process, adding no more than a few milliseconds of delay.
Meanwhile, vehicles that do not support the scheme will detect the signal as \SI{5}{\percent} within the allowed tolerance, and begin SLAC as normal.
Our proposed scheme is further illustrated in Figure~\ref{fig:new_proto} in the Appendix.

After a short period of this initial carrier, the charger would transmit a synchronization value to uniquely identify the start of the data, such as 0110, followed by the NMK bits, additional information and appropriate error correction.
At \SI{1}{\kilo \hertz}, the data rate is \SI{1}{\kilo bps}, sufficient to transfer the 128-bit NMK and some metadata in less than \SI{0.2}{\second}.
Vehicles which detected the signal would not start SLAC and simply wait to receive the NMK via this direct method.

Depending on how existing EV and charger hardware was made, it may be possible to implement our scheme using only a software update.
Modern micro-controllers contain versatile peripherals, that could easily generate such a signal with a different length for each pulse.
Similarly, sampling the signal at \SI{1}{MSps} would be enough to measure the duty cycle down to \SI{0.1}{\percent}.

\section{Discussion}

\subsection{Alternate Attack Hardware}

Due to the unique frequency range used by HomePlug, we used a USRP N210 for the analog-digital interface.
However, we did not take advantage of any of the advanced analog RF processing provided by this device, and a simple, high-bandwidth analog peripheral would be sufficient for our system.
For example, the low-cost Red Pitaya~\cite{redpitaya} hardware can sample and transmit at up to \SI{125}{\mega \hertz}.
It also has an on-board FPGA and Gigabit Ethernet, and therefore, it has all the necessary hardware to be used as an SDR for HomePlug, reducing the hardware cost from $\sim$\$3500 to $\sim$\$400.

Additionally, researchers have been successful in using off-the-shelf HomePlug modems to tamper with CCS.
As discussed, the hardware limitations of these devices makes it impossible to add additional RF amplifiers, greatly limiting their practical range, with most attacks requiring a direct connection to the data line, or a coil wrapped around the charging cable.
Additionally, the firmware of these modems prevents them from eavesdropping on all messages, requiring active techniques to redirect traffic.

\subsection{Further Attack Goals}

As CCS is a large and complex protocol, an attacker may have many different goals with their access to the communications.
These can range from the straightforward attacks against power delivery, to using it is an entry point deeper into the EV charging ecosystem.

\subsubsection{Load Alteration Attacks}
In addition to the slow charging and over-charging attacks evaluated in this paper, an attacker may wish to use EVs as an attack vector against the power infrastructure itself.
Load alteration attacks have been identified as a major emerging threat to the power grid as renewable energy ratios increase~\cite{soltan2018blackiot, shekari2022madiot}.
Modern EVs are a particularly attractive target for such an attack~\cite{ghafouri2022coordianted} due to their large potential power draw that can appear in an instant.
An attacker able to control power delivery on a fleet of high-powered chargers might effectively mount such an attack, by rapidly ramping the charging power up and down at just the right times.

\subsubsection{Plug and Charge Process}

An attacker may be motivated to perform billing fraud, either to enrich themselves or to cause financial harm to others.
The PnC process is an attractive target for such an attack, and researchers have proposed potential avenues for attack~\cite{conti2022evexchange, low2025security}.
Additionally, the attacker could aim to disable PnC, thereby greatly downgrading the user experience.

\subsubsection{Value Added Services}

In addition to the essential power and payment messages, CCS allows using the Ethernet link inside the charging cable for non-charging related services.
For example, the ``Internet'' service~\cite{std_ISO_15118_20} could allow a vehicle to connect to the internet through the charger, e.g. to download firmware updates.
This could grant the attacker access to many other sensitive data links coming out of the vehicles.
Novel services are also being proposed~\cite{silva2025evolve}, and an attacker could target any of them.

\subsubsection{Software Attacks}

Furthermore, since the communication link is a standard Ethernet connection, it is likely to be implemented as an Ethernet interface of the computer inside the device.
Additional services running on the device, e.g. management and debugging, can have open ports accessible via the cable~\cite{beijnum2025test}.
While an attacker can easily connect to a public CCS chargers without the need to MitM a connection, they cannot similarly access EVs, as the charging port is normally locked.
Additionally, MitM attacks offer the benefit of stealth without visibly plugging into the charger.

\section{Related Work}

Due to its importance, the security of the EV charging ecosystem has been studied for a long time~\cite{fries2012electric}.
Following its standardization, such work now also studies ISO 15118 specifically~\cite{bao2018threat, skarga-bandurova2022cyber}.

Prior work has created an SDR HPGP implementation~\cite{baker2019losing} that can eavesdrop on communications.
This implementation is not capable of real-time operation, it can only process pre-recorded RF data files.
SDRs have also been used to transmit the HomePlug preamble, denying service~\cite{kohler2023brokenwire}.
While this attack proved that the attacker can inject into the cable, the preamble is extremely easy to detect.
This left an open question as to whether data packets can also be successfully injected.

Without using an SDR, researchers have also managed to infiltrate the HPGP network, by using an off-the-shelf modem~\cite{dudek2019v2g}.
When correctly configured, modems are able to receive the \texttt{SLAC\_\allowbreak{}MATCH.\allowbreak{}CNF} messages, and therefore the NMK.
Timing attacks against SDP have been discussed~\cite{zhdanova2022local}, but the highly optimized SDP implementations within real EV chargers make this difficult and unreliable.
Motivated by this, researchers attacked the HomePlug modems directly by using special commands to overwrite the routing table~\cite{eder2025charging}.
This attack works because most HPGP modems allow their configuration to be overwritten, which was exploited more broadly for a permanent denial of service ``bricking'' attack~\cite{szakaly2025pibuster}.

MitM has also been achieved using an intermediate device with two HomePlug modems, which is inserted into the charging cable~\cite{chen2025case}.
While such a device allows access to the basic signaling lines and has benefits for laboratory testing, it does not demonstrate real-world attack capabilities.
In all cases, the internal Ethernet filtering of modems made it impossible to achieve full sniffing of the communications without actively routing data through the attackers MitM.
Our system allows flexibility by allowing for truly passive real-time eavesdropping.

Legacy (SAE J1772~\cite{std_SAE_J1772} / IEC 61851~\cite{std_IEC_61851_1}) AC charging is a very simple protocol.
However, it is much more widely used than DC and therefore it has also been studied.
Researchers have developed MitM devices that can be inserted into the charging cable, in order to modify the AC current limit or deny charging~\cite{javier2025mitm, shi2025physical, zhou2023chargex}.
These attacks target the basic signaling process, and therefore all require the attacker to insert a device into the charging cable.
We consider that such a modification is much more likely to be detected than our nearby wireless MitM.

Power systems have attracted much research, due to the ultimate risk of catastrophic damage.
Voltage sensors used within these devices can be wirelessly spoofed~\cite{dayanikli2020electromagnetic}, and therefore attackers can control the output voltage of power supplies~\cite{szakaly2024assault}.
Similar attacks have also been performed on inverters~\cite{dan2022novel}.
While it has been discussed, to our knowledge no-one has experimentally studied power delivery attacks on real EVs.

The PnC and payment system of CCS has also been considered as a potential target.
Researchers have identified multiple relay attacks on the V2G communications~\cite{conti2022evexchange, low2025security}, where a user with PnC could be tricked into paying for someone else's charging session.
Our MitM would be perfect at executing such attacks, by tapping into both charging cables.

\section{Conclusion}

In this paper we presented a detailed, experimentally tested evaluation of MitM attacks against the CCS charging communication.
We performed the first detailed real-world analysis of how these attacks can be executed wirelessly, on real hardware.
We developed and introduced a powerful SDR implementation of HomePlug Green PHY, allowing us to perform protocol attacks optimized for a wireless setting.
We presented a detailed evaluation of SDP implementations within real devices, and analyzed the timing characteristics of systems using large, real-world datasets.
Building on this, we proposed a variety of specific and experimentally proven techniques to hijack the connection.

We experimentally showed that the attacker can control or strip the use of TLS within the connection, and that they can similarly select the protocol version.
While the possibility of power delivery attacks has been discussed within the community, our work is the first to actually test and evaluate such attacks.
All in all, this work presented an in-depth analysis of attacks against all layers of the CCS protocol, clearly demonstrating the flaws within the protocol.
We hope that this work will motivate those in charge to make improvements as soon as possible.
To help with this, we enumerated the available countermeasures, and introduced a novel, downgrade-proof system that could ensure adequate encryption directly at the physical layer, preventing all other attacks.

\section*{Acknowledgments}

We would like to thank armasuisse Science + Technology for their support. 
Marcell was funded by the Engineering and Physical Sciences Research Council (EPSRC) and Sebastian was supported by the Royal Academy of Engineering and the Office of the Chief Science Adviser for National Security under the UK Intelligence Community Postdoctoral Research Fellowships programme.

\section*{Ethics Considerations}

Our findings were disclosed to CharIN, an industry association of EV charging stakeholders that leads the standardization and development of CCS, in November 2025.
This is being circulated among their members, and we engaged with various CharIN members to support the evaluation and mitigation of the presented issues.
Due to the fundamental nature of the attack, it is not possible for the issues to be fixed within any reasonable disclosure time frame.
While we hope that each party will individually deploy heuristics and hardening techniques as fast as possible, these will not fundamentally prevent attacks.

We believe that publishing our work as soon as possible, even before mitigations are in place, is in the best public interest.
The possibility of CCS MitM attacks have been discussed in security overview papers since 2018~\cite{bao2018threat}, however the underlying issues have not been addressed.
We believe that presenting a detailed evaluation and comprehensive control over the CCS protocol will convince the industry that this is not merely a theoretical, but a practical threat.
We hope that our findings will raise public awareness and motivate the industry to develop fixes as soon as possible.

However, our tool offers an easy to use, reliable, end-to-end system to perform wireless MitM attacks against CCS.
Therefore, we cannot ethically publish any part of this code, due to the potential for exploitation.
Having to recreate this system will dramatically raise the barrier of entry for such attacks, limiting the risks posed by motivated hobbyists.

\FloatBarrier

\bibliographystyle{ACM-Reference-Format}
\bibliography{references}


\begin{thebibliography}{46}


\ifx \showCODEN    \undefined \def \showCODEN     #1{\unskip}     \fi
\ifx \showDOI      \undefined \def \showDOI       #1{#1}\fi
\ifx \showISBNx    \undefined \def \showISBNx     #1{\unskip}     \fi
\ifx \showISBNxiii \undefined \def \showISBNxiii  #1{\unskip}     \fi
\ifx \showISSN     \undefined \def \showISSN      #1{\unskip}     \fi
\ifx \showLCCN     \undefined \def \showLCCN      #1{\unskip}     \fi
\ifx \shownote     \undefined \def \shownote      #1{#1}          \fi
\ifx \showarticletitle \undefined \def \showarticletitle #1{#1}   \fi
\ifx \showURL      \undefined \def \showURL       {\relax}        \fi
\providecommand\bibfield[2]{#2}
\providecommand\bibinfo[2]{#2}
\providecommand\natexlab[1]{#1}
\providecommand\showeprint[2][]{arXiv:#2}

\bibitem[Baker and Martinovic(2019)]%
        {baker2019losing}
\bibfield{author}{\bibinfo{person}{Richard Baker} {and} \bibinfo{person}{Ivan Martinovic}.} \bibinfo{year}{2019}\natexlab{}.
\newblock \showarticletitle{Losing the Car Keys: Wireless {PHY-Layer} Insecurity in {EV} Charging}. In \bibinfo{booktitle}{\emph{28th USENIX Security Symposium (USENIX Security 19)}}. \bibinfo{publisher}{USENIX Association}, \bibinfo{address}{Santa Clara, CA}, \bibinfo{pages}{407--424}.
\newblock
\showISBNx{978-1-939133-06-9}
\urldef\tempurl%
\url{https://www.usenix.org/conference/usenixsecurity19/presentation/baker}
\showURL{%
\tempurl}


\bibitem[Bao et~al\mbox{.}(2018)]%
        {bao2018threat}
\bibfield{author}{\bibinfo{person}{Kaibin Bao}, \bibinfo{person}{Hristo Valev}, \bibinfo{person}{Manuela Wagner}, {and} \bibinfo{person}{Hartmut Schmeck}.} \bibinfo{year}{2018}\natexlab{}.
\newblock \showarticletitle{A threat analysis of the vehicle-to-grid charging protocol {ISO} 15118}.
\newblock \bibinfo{journal}{\emph{Computer Science-Research and Development}} \bibinfo{volume}{33}, \bibinfo{number}{1-2} (\bibinfo{year}{2018}), \bibinfo{pages}{3--12}.
\newblock


\bibitem[{CharIN}(2015)]%
        {std_CCS}
\bibfield{author}{\bibinfo{person}{{CharIN}}.} \bibinfo{year}{2015}\natexlab{}.
\newblock \bibinfo{booktitle}{\emph{Combined Charging System 1.0 Specification - {CCS} 1.0}}.
\newblock \bibinfo{type}{{T}echnical {R}eport}. \bibinfo{institution}{Coordination Office Charging Interface, c/o Carmeq GmbH}.
\newblock


\bibitem[{CharIN}(2022)]%
        {std_MCSv1}
\bibfield{author}{\bibinfo{person}{{CharIN}}.} \bibinfo{year}{2022}\natexlab{}.
\newblock \bibinfo{title}{{CharIN} Whitepaper Megawatt Charging System ({MCS}) v1.0}.
\newblock \bibinfo{howpublished}{\url{https://www.charin.global/media/pages/technology/knowledge-base/c708ba3361-1670238823/whitepaper_megawatt_charging_system_1.0.pdf}}.
\newblock
\newblock
\shownote{Accessed 2026-01-11}.


\bibitem[{CharIN}(2025)]%
        {std_MCS}
\bibfield{author}{\bibinfo{person}{{CharIN}}.} \bibinfo{year}{2025}\natexlab{}.
\newblock \bibinfo{title}{{CharIN} Whitepaper Megawatt Charging System ({MCS}) v2.0}.
\newblock \bibinfo{howpublished}{\url{https://www.charin.global/media/pages/technology/knowledge-base/0c2cc2c8da-1747654352/250508_whitepaper_megawatt_charging_system_2.0.pdf}}.
\newblock
\newblock
\shownote{Accessed 2026-01-11}.


\bibitem[Chen et~al\mbox{.}(2025)]%
        {chen2025case}
\bibfield{author}{\bibinfo{person}{Kuan-Yu Chen}, \bibinfo{person}{Xin-Xian Lin}, \bibinfo{person}{Chung-You Chen}, \bibinfo{person}{Wen~Wei Li}, {and} \bibinfo{person}{Shi-Cho Cha}.} \bibinfo{year}{2025}\natexlab{}.
\newblock \showarticletitle{A Case Study on Real-World MITM Attacks Against ISO 15118-2 and DIN 70121}. In \bibinfo{booktitle}{\emph{2025 IEEE 14th Global Conference on Consumer Electronics (GCCE)}}. \bibinfo{publisher}{Institute of Electrical and Electronics Engineers}, \bibinfo{pages}{595--598}.
\newblock
\urldef\tempurl%
\url{https://doi.org/10.1109/GCCE65946.2025.11274985}
\showDOI{\tempurl}


\bibitem[Conti et~al\mbox{.}(2022)]%
        {conti2022evexchange}
\bibfield{author}{\bibinfo{person}{Mauro Conti}, \bibinfo{person}{Denis Donadel}, \bibinfo{person}{Radha Poovendran}, {and} \bibinfo{person}{Federico Turrin}.} \bibinfo{year}{2022}\natexlab{}.
\newblock \showarticletitle{{EVE}xchange: A Relay Attack on Electric Vehicle Charging System}. In \bibinfo{booktitle}{\emph{Computer Security -- ESORICS 2022}}, \bibfield{editor}{\bibinfo{person}{Vijayalakshmi Atluri}, \bibinfo{person}{Roberto Di~Pietro}, \bibinfo{person}{Christian~D. Jensen}, {and} \bibinfo{person}{Weizhi Meng}} (Eds.). \bibinfo{publisher}{Springer International Publishing}, \bibinfo{address}{Cham}, \bibinfo{pages}{488--508}.
\newblock
\showISBNx{978-3-031-17140-6}


\bibitem[Dan et~al\mbox{.}(2022)]%
        {dan2022novel}
\bibfield{author}{\bibinfo{person}{Zihao Dan}, \bibinfo{person}{Fengchen Yang}, \bibinfo{person}{Kaikai Pan}, \bibinfo{person}{Chen Yan}, \bibinfo{person}{Xiaoyu Ji}, {and} \bibinfo{person}{Wenyuan Xu}.} \bibinfo{year}{2022}\natexlab{}.
\newblock \showarticletitle{A Novel EMI Attack Exploiting the Control Vulnerability of Photovoltaic Inverters}. In \bibinfo{booktitle}{\emph{2022 IEEE 6th Conference on Energy Internet and Energy System Integration (EI2)}}. \bibinfo{publisher}{Institute of Electrical and Electronics Engineers}, \bibinfo{pages}{1076--1080}.
\newblock
\urldef\tempurl%
\url{https://doi.org/10.1109/EI256261.2022.10116254}
\showDOI{\tempurl}


\bibitem[Dayanikli et~al\mbox{.}(2020)]%
        {dayanikli2020electromagnetic}
\bibfield{author}{\bibinfo{person}{G{\"o}k{\c{c}}en~Yilmaz Dayanikli}, \bibinfo{person}{Rees~R Hatch}, \bibinfo{person}{Ryan~M Gerdes}, \bibinfo{person}{Hongjie Wang}, {and} \bibinfo{person}{Regan Zane}.} \bibinfo{year}{2020}\natexlab{}.
\newblock \showarticletitle{Electromagnetic Sensor and Actuator Attacks on Power Converters for Electric Vehicles}. In \bibinfo{booktitle}{\emph{2020 IEEE Security and Privacy Workshops (SPW)}}. IEEE, \bibinfo{pages}{98--103}.
\newblock


\bibitem[Dudek et~al\mbox{.}(2019)]%
        {dudek2019v2g}
\bibfield{author}{\bibinfo{person}{Sébastien Dudek}, \bibinfo{person}{Jean-Christophe Delaunay}, {and} \bibinfo{person}{Vincent Fargues}.} \bibinfo{year}{2019}\natexlab{}.
\newblock \showarticletitle{{V2G} Injector: Whispering to cars and charging units through the Power-Line}. In \bibinfo{booktitle}{\emph{Proceedings of the SSTIC (Symposium sur la sécurité des technologies de l'information et des communications), Rennes, France}}.
\newblock


\bibitem[Eder et~al\mbox{.}(2025)]%
        {eder2025charging}
\bibfield{author}{\bibinfo{person}{Lukas Eder}, \bibinfo{person}{Jakob L\"{o}w}, {and} \bibinfo{person}{Hans-Joachim Hof}.} \bibinfo{year}{2025}\natexlab{}.
\newblock \showarticletitle{Charging Communication Sniffing and Man-in-the-Middle Attacks}. In \bibinfo{booktitle}{\emph{Proceedings of the 16th ACM International Conference on Future and Sustainable Energy Systems}} \emph{(\bibinfo{series}{E-Energy '25})}. \bibinfo{publisher}{Association for Computing Machinery}, \bibinfo{address}{New York, NY, USA}, \bibinfo{pages}{799–804}.
\newblock
\showISBNx{9798400711251}
\urldef\tempurl%
\url{https://doi.org/10.1145/3679240.3734648}
\showDOI{\tempurl}


\bibitem[{European Commission}(2014)]%
        {ccs_law}
\bibfield{author}{\bibinfo{person}{{European Commission}}.} \bibinfo{year}{2014}\natexlab{}.
\newblock \showarticletitle{DIRECTIVE 2014/94/{EU} OF THE EUROPEAN PARLIAMENT AND OF THE COUNCIL}.
\newblock  (\bibinfo{date}{22 10} \bibinfo{year}{2014}).
\newblock
\urldef\tempurl%
\url{https://eur-lex.europa.eu/legal-content/EN/TXT/HTML/?uri=CELEX:32014L0094&from=en}
\showURL{%
\tempurl}


\bibitem[{European Commission}(2025)]%
        {15118_law}
\bibfield{author}{\bibinfo{person}{{European Commission}}.} \bibinfo{year}{2025}\natexlab{}.
\newblock \bibinfo{title}{{Commission Delegated Regulation (EU) 2025/656 of 2 April 2025 amending Regulation (EU) 2023/1804 of the European Parliament and of the Council as regards standards for wireless recharging, electric road system, vehicle-to-grid communication and hydrogen supply for road transport vehicles}}.
\newblock
\newblock
\newblock
\shownote{\\\url{https://eur-lex.europa.eu/legal-content/EN/TXT/?uri=CELEX:32025R0656}}.


\bibitem[{Federal Highway Administration}(2023)]%
        {us_nevi}
\bibfield{author}{\bibinfo{person}{{Federal Highway Administration}}.} \bibinfo{year}{2023}\natexlab{}.
\newblock \bibinfo{title}{{88 FR 12724}, National Electric Vehicle Infrastructure Standards and Requirements}.
\newblock \bibinfo{howpublished}{\url{https://www.federalregister.gov/d/2023-03500}}.
\newblock


\bibitem[Fleck and Dimov(2001)]%
        {fleck2001wireless}
\bibfield{author}{\bibinfo{person}{Bob Fleck} {and} \bibinfo{person}{Jordan Dimov}.} \bibinfo{year}{2001}\natexlab{}.
\newblock \bibinfo{title}{Wireless Access Points and {ARP} Poisoning}.
\newblock \bibinfo{howpublished}{http://target0.be/madchat/reseau/wireless/ARP\_Poisoning/wirelessAcessPoints-ARPPoisoning.pdf}.
\newblock
\newblock
\shownote{Accessed 2026-01-14}.


\bibitem[Fries and Falk(2012)]%
        {fries2012electric}
\bibfield{author}{\bibinfo{person}{Steffen Fries} {and} \bibinfo{person}{Rainer Falk}.} \bibinfo{year}{2012}\natexlab{}.
\newblock \showarticletitle{Electric Vehicle Charging Infrastructure - Security Considerations and Approaches}.
\newblock


\bibitem[Ghafouri et~al\mbox{.}(2022)]%
        {ghafouri2022coordianted}
\bibfield{author}{\bibinfo{person}{Mohsen Ghafouri}, \bibinfo{person}{Ekram Kabir}, \bibinfo{person}{Bassam Moussa}, {and} \bibinfo{person}{Chadi Assi}.} \bibinfo{year}{2022}\natexlab{}.
\newblock \showarticletitle{Coordinated Charging and Discharging of Electric Vehicles: A New Class of Switching Attacks}.
\newblock \bibinfo{journal}{\emph{ACM Trans. Cyber-Phys. Syst.}} \bibinfo{volume}{6}, \bibinfo{number}{3}, Article \bibinfo{articleno}{23} (\bibinfo{date}{Sept.} \bibinfo{year}{2022}), \bibinfo{numpages}{26}~pages.
\newblock
\showISSN{2378-962X}
\urldef\tempurl%
\url{https://doi.org/10.1145/3524454}
\showDOI{\tempurl}


\bibitem[Harper(2013)]%
        {harper2013development}
\bibfield{author}{\bibinfo{person}{Jason~D Harper}.} \bibinfo{year}{2013}\natexlab{}.
\newblock \showarticletitle{Development and Implementation of SAE DC Charging Digital Communication for Plug-in Electric Vehicle DC Charging}.
\newblock \bibinfo{journal}{\emph{SAE International}} (\bibinfo{year}{2013}).
\newblock
\showISSN{0148-7191}


\bibitem[{HomePlug Powerline Alliance}(2013)]%
        {std_HPGP}
\bibfield{author}{\bibinfo{person}{{HomePlug Powerline Alliance}}.} \bibinfo{year}{2013}\natexlab{}.
\newblock \bibinfo{title}{{H}ome{P}lug Green {PHY} Specification, Release Version 1.1.1}.
\newblock
\newblock
\newblock
\shownote{\\\url{https://web.archive.org/web/20180825120357/https://www.homeplug.org/media/filer_public/74/40/7440ccd5-8c66-49ed-a2ce-5ef661932c27/homeplug_gp_specification_v111_final_public.pdf}}.


\bibitem[{International Electrotechnical Commission}(2017)]%
        {std_IEC_61851_1}
\bibfield{author}{\bibinfo{person}{{International Electrotechnical Commission}}.} \bibinfo{year}{2017}\natexlab{}.
\newblock \bibinfo{title}{{IEC} 61851-1. Electric vehicle conductive charging system. Part 1. General requirements}.
\newblock
\newblock


\bibitem[{International Organization for Standardization}(2014)]%
        {std_ISO_15118_2}
\bibfield{author}{\bibinfo{person}{{International Organization for Standardization}}.} \bibinfo{year}{2014}\natexlab{}.
\newblock \bibinfo{title}{{ISO} 15118-2. Vehicle to grid communication interface. Part 2: Network and application protocol requirements}.
\newblock
\newblock


\bibitem[{International Organization for Standardization}(2019)]%
        {std_ISO_15118_1}
\bibfield{author}{\bibinfo{person}{{International Organization for Standardization}}.} \bibinfo{year}{2019}\natexlab{}.
\newblock \bibinfo{title}{{ISO} 15118-1. Vehicle to grid communication interface. Part 1: General information and use-case definition}.
\newblock
\newblock


\bibitem[{International Organization for Standardization}(2022)]%
        {std_ISO_15118_20}
\bibfield{author}{\bibinfo{person}{{International Organization for Standardization}}.} \bibinfo{year}{2022}\natexlab{}.
\newblock \bibinfo{title}{{ISO} 15118-20. Vehicle to grid communication interface. Part 20: 2nd generation network layer and application layer requirements}.
\newblock
\newblock


\bibitem[Jarauta~Gastelu et~al\mbox{.}(2025)]%
        {javier2025mitm}
\bibfield{author}{\bibinfo{person}{Javier Jarauta~Gastelu}, \bibinfo{person}{Roberto Gesteira-Miñarro}, \bibinfo{person}{Javier Matanza}, \bibinfo{person}{Rafael Palacios}, {and} \bibinfo{person}{Gregorio López}.} \bibinfo{year}{2025}\natexlab{}.
\newblock \showarticletitle{MitM Attack to Electric Vehicle AC Chargers}.
\newblock \bibinfo{journal}{\emph{IEEE Internet of Things Journal}} \bibinfo{volume}{12}, \bibinfo{number}{19} (\bibinfo{year}{2025}), \bibinfo{pages}{39689--39700}.
\newblock
\urldef\tempurl%
\url{https://doi.org/10.1109/JIOT.2025.3589219}
\showDOI{\tempurl}


\bibitem[Köhler et~al\mbox{.}(2023)]%
        {kohler2023brokenwire}
\bibfield{author}{\bibinfo{person}{Sebastian Köhler}, \bibinfo{person}{Richard Baker}, \bibinfo{person}{Martin Strohmeier}, {and} \bibinfo{person}{Ivan Martinovic}.} \bibinfo{year}{2023}\natexlab{}.
\newblock \showarticletitle{Brokenwire : Wireless Disruption of {CCS} Electric Vehicle Charging}. In \bibinfo{booktitle}{\emph{NDSS Symposium}}.
\newblock


\bibitem[Lampe et~al\mbox{.}(2016)]%
        {lampe2016power}
\bibfield{author}{\bibinfo{person}{Lutz Lampe}, \bibinfo{person}{Andrea~M. Tonello}, {and} \bibinfo{person}{Theo~G. Swart}.} \bibinfo{year}{2016}\natexlab{}.
\newblock \bibinfo{booktitle}{\emph{Power Line Communications: Principles, Standards and Applications from Multimedia to Smart Grid} (\bibinfo{edition}{2nd} ed.)}.
\newblock \bibinfo{publisher}{Wiley Publishing}.
\newblock
\showISBNx{1118676718, 9781118676714}


\bibitem[Löw et~al\mbox{.}(2025a)]%
        {low2025draindead}
\bibfield{author}{\bibinfo{person}{Jakob Löw}, \bibinfo{person}{Dominik Bayerl}, \bibinfo{person}{Kevin Mayer}, {and} \bibinfo{person}{Hans-Joachim Hof}.} \bibinfo{year}{2025}\natexlab{a}.
\newblock \showarticletitle{DrainDead: Emptying Batteries of Parked Electric Vehicles}. In \bibinfo{booktitle}{\emph{Proceedings of the 3rd USENIX Symposium on Vehicle Security \& Privacy (VehicleSec ’25)}}. \bibinfo{publisher}{USENIX Association}.
\newblock
\urldef\tempurl%
\url{https://www.usenix.org/system/files/vehiclesec25-low.pdf}
\showURL{%
\tempurl}


\bibitem[Löw et~al\mbox{.}(2025b)]%
        {low2025security}
\bibfield{author}{\bibinfo{person}{Jakob Löw}, \bibinfo{person}{Vishwa Vasu}, \bibinfo{person}{Thomas Hutzelmann}, {and} \bibinfo{person}{Hans-Joachim Hof}.} \bibinfo{year}{2025}\natexlab{b}.
\newblock \bibinfo{title}{Security Aspects of ISO 15118 Plug and Charge Payment}.
\newblock
\newblock
\showeprint[arxiv]{2512.15966}~[cs.CR]
\urldef\tempurl%
\url{https://arxiv.org/abs/2512.15966}
\showURL{%
\tempurl}


\bibitem[Molliere(2024)]%
        {ev_bus}
\bibfield{author}{\bibinfo{person}{Max Molliere}.} \bibinfo{year}{2024}\natexlab{}.
\newblock \showarticletitle{Battery-electric is now the most popular for new city buses in the EU}.
\newblock \bibinfo{howpublished}{\url{https://www.transportenvironment.org/articles/battery-electric-is-now-the-top-powertrain-type-for-new-city-buses-in-the-eu}}.
\newblock \bibinfo{journal}{\emph{Transport \& Environment}} (\bibinfo{date}{05 07} \bibinfo{year}{2024}).
\newblock
\newblock
\shownote{Accessed 2026-01-14}.


\bibitem[{NHS England}(2022)]%
        {ev_fleet_nhs}
\bibfield{author}{\bibinfo{person}{{NHS England}}.} \bibinfo{year}{2022}\natexlab{}.
\newblock \bibinfo{title}{{NHS} rolls out new electric vehicles to help patients and the planet}.
\newblock \bibinfo{howpublished}{\\\url{https://www.england.nhs.uk/2022/08/nhs-rolls-out-new-electric-vehicles-to-help-patients-and-the-planet/}}.
\newblock
\newblock
\shownote{Accessed 2026-01-14}.


\bibitem[{Red Pitaya}(2025)]%
        {redpitaya}
\bibfield{author}{\bibinfo{person}{{Red Pitaya}}.} \bibinfo{year}{2025}\natexlab{}.
\newblock \bibinfo{title}{STEMlab 125-14}.
\newblock \bibinfo{howpublished}{\url{https://redpitaya.com/stemlab-125-14/}}.
\newblock
\newblock
\shownote{Accessed: 2026-01-14}.


\bibitem[{SAE International}(1996)]%
        {std_SAE_J1772}
\bibfield{author}{\bibinfo{person}{{SAE International}}.} \bibinfo{year}{1996}\natexlab{}.
\newblock \bibinfo{title}{{SAE} J1772: Electric Vehicle and Plug in Hybrid Electric Vehicle Conductive Charge Coupler}.
\newblock
\newblock


\bibitem[Shekari et~al\mbox{.}(2022)]%
        {shekari2022madiot}
\bibfield{author}{\bibinfo{person}{Tohid Shekari}, \bibinfo{person}{Alvaro~A. Cardenas}, {and} \bibinfo{person}{Raheem Beyah}.} \bibinfo{year}{2022}\natexlab{}.
\newblock \showarticletitle{{MaDIoT} 2.0: Modern {High-Wattage} {IoT} Botnet Attacks and Defenses}. In \bibinfo{booktitle}{\emph{31st USENIX Security Symposium (USENIX Security 22)}}. \bibinfo{publisher}{USENIX Association}, \bibinfo{address}{Boston, MA}, \bibinfo{pages}{3539--3556}.
\newblock
\showISBNx{978-1-939133-31-1}
\urldef\tempurl%
\url{https://www.usenix.org/conference/usenixsecurity22/presentation/shekari}
\showURL{%
\tempurl}


\bibitem[Shi et~al\mbox{.}(2025)]%
        {shi2025physical}
\bibfield{author}{\bibinfo{person}{Hetian Shi}, \bibinfo{person}{Yi He}, \bibinfo{person}{Shangru Song}, \bibinfo{person}{Jianwei Zhuge}, {and} \bibinfo{person}{Jian Mao}.} \bibinfo{year}{2025}\natexlab{}.
\newblock \bibinfo{title}{Physical-Layer Signal Injection Attacks on EV Charging Ports: Bypassing Authentication via Electrical-Level Exploits}.
\newblock
\newblock
\showeprint[arxiv]{2506.16400}~[cs.CR]
\urldef\tempurl%
\url{https://arxiv.org/abs/2506.16400}
\showURL{%
\tempurl}


\bibitem[Silva et~al\mbox{.}(2025)]%
        {silva2025evolve}
\bibfield{author}{\bibinfo{person}{Erick Silva}, \bibinfo{person}{Tadeu Freitas}, \bibinfo{person}{Rehana Yasmin}, \bibinfo{person}{Ali Shoker}, {and} \bibinfo{person}{Paulo Esteves-Verissimo}.} \bibinfo{year}{2025}\natexlab{}.
\newblock \showarticletitle{EVolve: A Value-Added Services Platform for Electric Vehicle Charging Stations}. In \bibinfo{booktitle}{\emph{2025 IEEE 101st Vehicular Technology Conference (VTC2025-Spring)}}. \bibinfo{pages}{1--7}.
\newblock
\urldef\tempurl%
\url{https://doi.org/10.1109/VTC2025-Spring65109.2025.11174448}
\showDOI{\tempurl}


\bibitem[Simpson et~al\mbox{.}(2007)]%
        {rfc4861}
\bibfield{author}{\bibinfo{person}{William~A. Simpson}, \bibinfo{person}{Dr.~Thomas Narten}, \bibinfo{person}{Erik Nordmark}, {and} \bibinfo{person}{Hesham Soliman}.} \bibinfo{year}{2007}\natexlab{}.
\newblock \bibinfo{title}{{Neighbor Discovery for IP version 6 (IPv6)}}.
\newblock \bibinfo{howpublished}{RFC 4861}.
\newblock
\urldef\tempurl%
\url{https://doi.org/10.17487/RFC4861}
\showDOI{\tempurl}


\bibitem[Skarga-Bandurova et~al\mbox{.}(2022)]%
        {skarga-bandurova2022cyber}
\bibfield{author}{\bibinfo{person}{Inna Skarga-Bandurova}, \bibinfo{person}{Igor Kotsiuba}, {and} \bibinfo{person}{Tetiana Biloborodova}.} \bibinfo{year}{2022}\natexlab{}.
\newblock \showarticletitle{Cyber Security of Electric Vehicle Charging Infrastructure: Open Issues and Recommendations}. In \bibinfo{booktitle}{\emph{2022 IEEE International Conference on Big Data (Big Data)}}. \bibinfo{pages}{3099--3106}.
\newblock
\urldef\tempurl%
\url{https://doi.org/10.1109/BigData55660.2022.10020644}
\showDOI{\tempurl}


\bibitem[Soltan et~al\mbox{.}(2018)]%
        {soltan2018blackiot}
\bibfield{author}{\bibinfo{person}{Saleh Soltan}, \bibinfo{person}{Prateek Mittal}, {and} \bibinfo{person}{H~Vincent Poor}.} \bibinfo{year}{2018}\natexlab{}.
\newblock \showarticletitle{{BlackIoT:IoT Botnet of High Wattage Devices Can Disrupt the Power Grid}}. In \bibinfo{booktitle}{\emph{27th USENIX Security Symposium (USENIX Security 18)}}. \bibinfo{pages}{15--32}.
\newblock


\bibitem[Szak{\'a}ly et~al\mbox{.}(2025)]%
        {szakaly2025pibuster}
\bibfield{author}{\bibinfo{person}{Marcell Szak{\'a}ly}, \bibinfo{person}{Sebastian K{\"o}hler}, {and} \bibinfo{person}{Ivan Martinovic}.} \bibinfo{year}{2025}\natexlab{}.
\newblock \showarticletitle{Short: {PIB}uster: Exploiting a Common Misconfiguration in {CCS} {EV} Chargers}. In \bibinfo{booktitle}{\emph{3rd USENIX Symposium on Vehicle Security and Privacy (VehicleSec '25)}}.
\newblock


\bibitem[Szakály et~al\mbox{.}(2025)]%
        {szakaly2025current}
\bibfield{author}{\bibinfo{person}{Marcell Szakály}, \bibinfo{person}{Sebastian K{\"o}hler}, {and} \bibinfo{person}{Ivan Martinovic}.} \bibinfo{year}{2025}\natexlab{}.
\newblock \showarticletitle{Current Affairs: A Security Measurement Study of {CCS} {EV} Charging Deployments}. In \bibinfo{booktitle}{\emph{34th USENIX Security Symposium}}.
\newblock


\bibitem[Szakály et~al\mbox{.}(2024)]%
        {szakaly2024assault}
\bibfield{author}{\bibinfo{person}{Marcell Szakály}, \bibinfo{person}{Sebastian Köhler}, \bibinfo{person}{Martin Strohmeier}, {and} \bibinfo{person}{Ivan Martinovic}.} \bibinfo{year}{2024}\natexlab{}.
\newblock \showarticletitle{Assault and Battery: Evaluating the Security of Power Conversion Systems Against Electromagnetic Injection Attacks}. In \bibinfo{booktitle}{\emph{2024 Annual Computer Security Applications Conference (ACSAC)}}. \bibinfo{publisher}{IEEE Computer Society}, \bibinfo{pages}{224--239}.
\newblock
\urldef\tempurl%
\url{https://doi.org/10.1109/ACSAC63791.2024.00033}
\showDOI{\tempurl}


\bibitem[{Tesla}(2022)]%
        {std_NACS}
\bibfield{author}{\bibinfo{person}{{Tesla}}.} \bibinfo{year}{2022}\natexlab{}.
\newblock \bibinfo{title}{TS-0023666: North American Charging Standard Technical Specification}.
\newblock
\newblock


\bibitem[{UPS}(2020)]%
        {ev_fleet_ups}
\bibfield{author}{\bibinfo{person}{{UPS}}.} \bibinfo{year}{2020}\natexlab{}.
\newblock \bibinfo{title}{{UPS} invests in Arrival, accelerates fleet electrification with a commitment to purchase up to 10,000 electric vehicles}.
\newblock \bibinfo{howpublished}{\\\url{https://about.ups.com/ca/en/newsroom/press-releases/sustainable-services/ups-invests-in-arrival-accelerates-fleet-electrification-with-order-of-10-000-electric-delivery-vehicles.html}}.
\newblock
\newblock
\shownote{Accessed 2026-01-16}.


\bibitem[{van Beijnum} et~al\mbox{.}(2025)]%
        {beijnum2025test}
\bibfield{author}{\bibinfo{person}{Wilco {van Beijnum}}, \bibinfo{person}{Harm van~den Brink}, {and} \bibinfo{person}{Paul Broos}.} \bibinfo{year}{2025}\natexlab{}.
\newblock \bibinfo{booktitle}{\emph{Test Report on Cyber Security Issues {AC} and {DC} Charging Stations}}.
\newblock \bibinfo{type}{{T}echnical {R}eport}. \bibinfo{institution}{ElaadNL}.
\newblock
\newblock
\shownote{Accessed 2026-01-14}.


\bibitem[Zhdanova et~al\mbox{.}(2022)]%
        {zhdanova2022local}
\bibfield{author}{\bibinfo{person}{Maria Zhdanova}, \bibinfo{person}{Julian Urbansky}, \bibinfo{person}{Anne Hagemeier}, \bibinfo{person}{Daniel Zelle}, \bibinfo{person}{Isabelle Herrmann}, {and} \bibinfo{person}{Dorian H\"{o}ffner}.} \bibinfo{year}{2022}\natexlab{}.
\newblock \showarticletitle{Local Power Grids at Risk - An Experimental and Simulation-based Analysis of Attacks on Vehicle-To-Grid Communication}. In \bibinfo{booktitle}{\emph{Proceedings of the 38th Annual Computer Security Applications Conference}} (Austin, TX, USA) \emph{(\bibinfo{series}{ACSAC '22})}. \bibinfo{publisher}{Association for Computing Machinery}, \bibinfo{address}{New York, NY, USA}, \bibinfo{pages}{42--55}.
\newblock
\showISBNx{9781450397599}
\urldef\tempurl%
\url{https://doi.org/10.1145/3564625.3568136}
\showDOI{\tempurl}


\bibitem[Zhou et~al\mbox{.}(2023)]%
        {zhou2023chargex}
\bibfield{author}{\bibinfo{person}{Ce Zhou}, \bibinfo{person}{Qiben Yan}, \bibinfo{person}{Zhiyuan Yu}, \bibinfo{person}{Eshan Dixit}, \bibinfo{person}{Ning Zhang}, \bibinfo{person}{Huacheng Zeng}, {and} \bibinfo{person}{Alireza~Safdari Ghanhdari}.} \bibinfo{year}{2023}\natexlab{}.
\newblock \bibinfo{title}{{C}harge{X}: Exploring State Switching Attack on Electric Vehicle Charging Systems}.
\newblock \bibinfo{howpublished}{\url{https://arxiv.org/abs/2305.08037}}.
\newblock
\showeprint[arxiv]{2305.08037}~[cs.CR]


\end{thebibliography}

\appendix

\section{Attack Tool Description}
\label{app:tool}

In this section we present additional information about the implementation of our SDR and MitM tools.

\begin{figure}[t]
    \centering
    \includegraphics[width=0.95\linewidth]{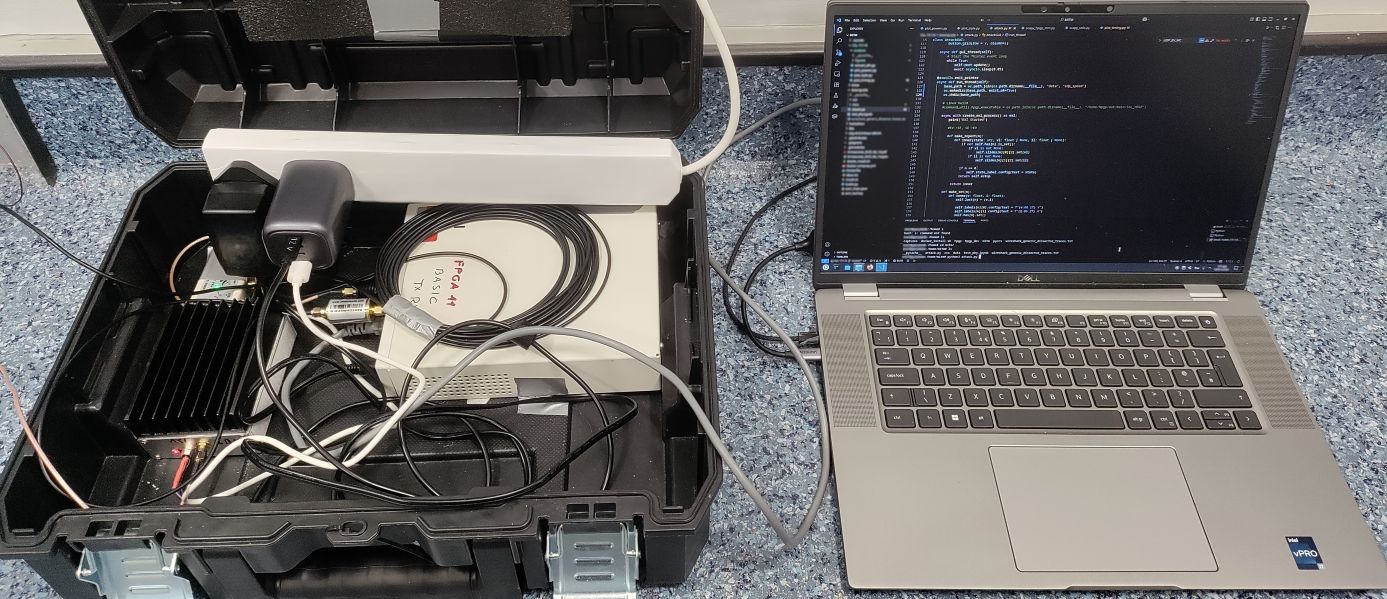}
    \caption{The MitM hardware. To the left, a transport case with with the TX amplifier (left), RX amplifier (top left), and USRP (right). On the right, the laptop used to run the software. The two cables going to the left connect to the antennas.}
    \Description{A black transport case containing RF hardware, and a laptop with code on the screen.}
    \label{fig:box}
\end{figure}

\paragraph{Hardware}
The attack hardware can be seen in Figure~\ref{fig:box}.
The necessary equipment has been mounted in a transport case, and includes the USRP, a TX amplifier, an RX amplifier, and power supplies.
To improve reception quality, we also added a Mini Circuits SLP-36+, DC-\SI{36}{\mega \hertz} low-pass filter between the RX amplifier and the USRP, to filter out any higher frequency noise.

\begin{figure}[t]
    \centering
    \includegraphics[width=0.95\linewidth]{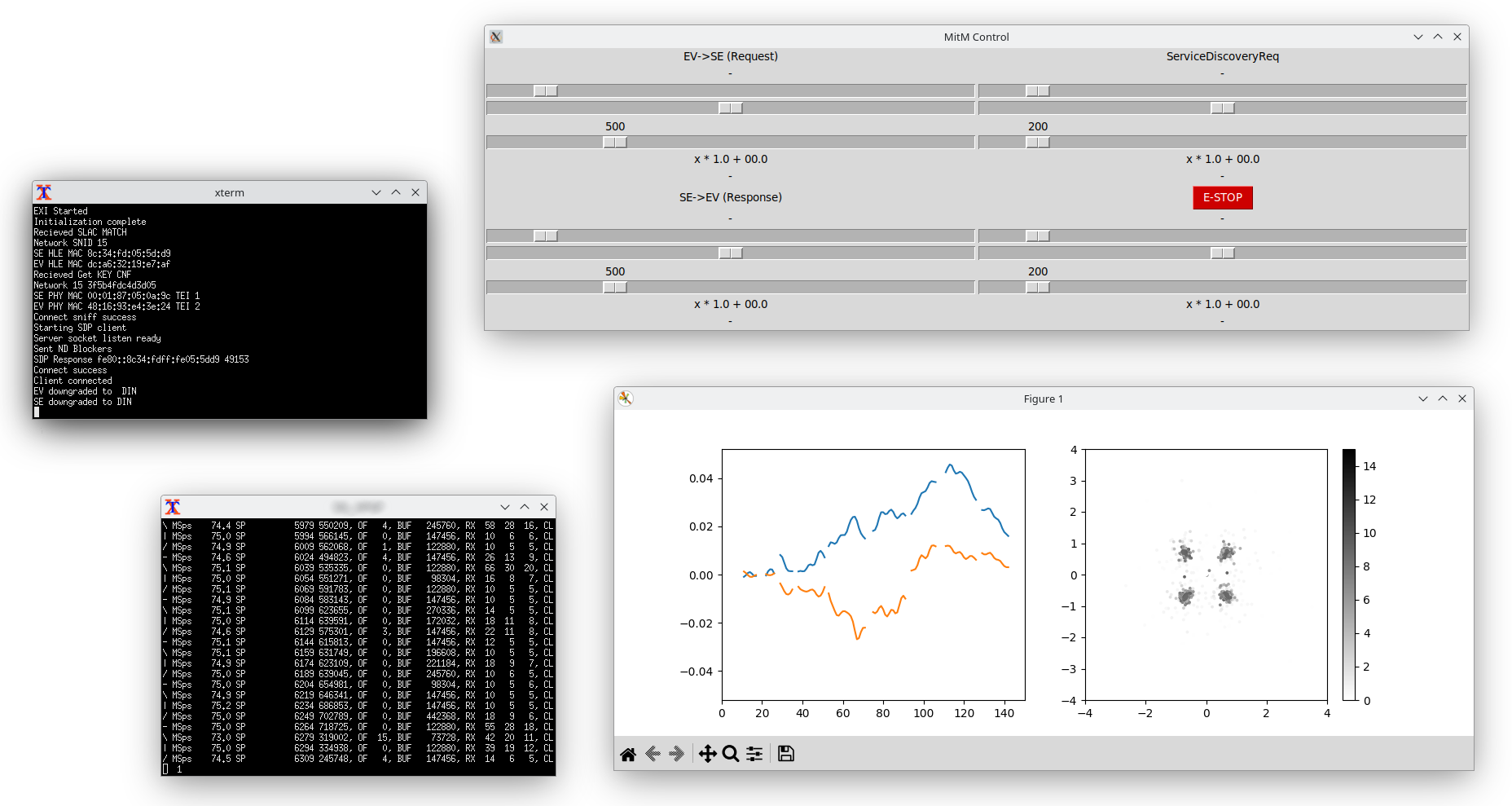}
    \caption{The user interface for MitM attacks. Top left: status messages from the MitM tool. Top right: Power delivery editing panel. The current protocol state is shown in the top right. Once the charging enters power delivery, the sliders can be used to change parameters. Bottom left: Performance monitoring from the SDR. Bottom right: Live view of the analog signal quality.}
    \Description{}
    \label{fig:attack_screen}
\end{figure}

\paragraph{SDR Driver}
Our software consists of two separate parts.
First, the SDR driver is responsible for converting between the analog waveform and Ethernet frames.
This is implemented entirely within C++, using to the USRP UHD drivers to access the hardware.
The code performs digital signal processing calculations on the incoming data stream to identify HPGP packets, correct signal distortions, and extract the modulated data.
In HPGP, data is carried using Quadrature Phase Shift Keying, which causes received data points to form a square if processed correctly.
The complex valued transfer function of the wireless channel across the sub-carriers, as well as the constellation diagram showing the square are shown to the user as seen in Figure~\ref{fig:attack_screen}.

The BCJR algorithm is used to perform soft decoding of the Turbo error correction.
The bits are processed according to the HPGP standard, checksums are validated, and data is decrypted wherever possible.
The segmented data frames are re-assembled into Ethernet frames, and these are provided for further processing by other applications.
Additionally, the implementation automatically extracts HPGP encryption keys found in any management frame, to aid with further decryption.

When transmitting, the same process is performed in reverse.
The SDR knows which devices it is transmitting to, and automatically selects encryption keys that they already know.
The data is segmented, encrypted, error corrected, modulated, and sent to the hardware.

\paragraph{MitM Utility}
For flexibility, the MitM utility is implemented in Python.
It connects to the SDR driver via a local socket, to get access to the SDR data stream.
Within the MitM, packets are parsed and processed using Scapy.
The tool contains a custom implementation of UDP, SDP, as well as the TCP stack.
This allows us to process packets which would usually dropped by the IP stack, for example due to destination address filtering.
Additionally, using a custom TCP implementation allows fine-grained control over re-transmissions, improving the reliability of our system during packet loss.

\begin{figure}[t]
    \centering
    \includegraphics[width=0.95\linewidth]{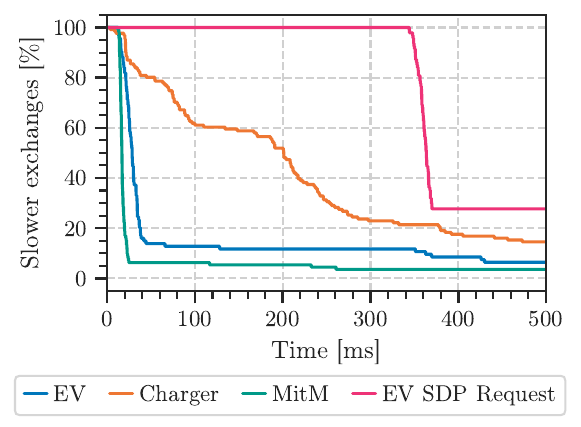}
    \caption{Comparison of the latency of EV, Charger, and MitM HPGP implementations in sending the first data packets after the encryption keys were shared.}
    \Description{}
    \label{fig:join_speed}
\end{figure}

The attacker is presented with a panel of values and sliders as shown in Figure~\ref{fig:attack_screen}.
In total, there are 4 groups of sliders corresponding to the voltage and current values, going in both directions.
The panel shows the latest values received and being sent, and the sliders can be used to change the rewriting rules.
Values are re-written using a linear relationship from two sliders $a, b$ using $x \rightarrow ax+b$.
This allows many useful tests, including offsets, multipliers, or replacing with a constant.
There is also an emergency stop button which replaces all future current requests going to the charger to ask for \SI{0}{\volt}, \SI{0}{\ampere}, and requests to immediately terminate the charging session.

\section{Further Modem Timing Analysis}
\label{app:join_speed}

\begin{figure}[t]
    \centering
    \includegraphics[width=0.9\linewidth,trim={0cm 4cm 0cm 0cm},clip]{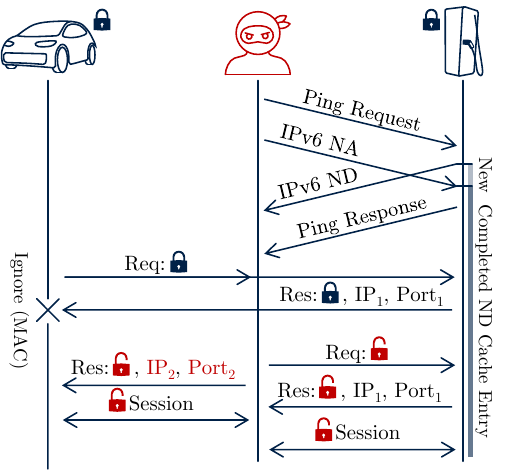}
    \caption{The NA Spoofing Attack. The attacker uses a ping to create the ND Cache entry, and immediately fills it with a spoofed ND response. This causes the SDP response to be dropped by the EV due to an invalid destination MAC address. The attacker can then hijack SDP without a strict time constraint.}
    \Description{}
    \label{fig:na_attack}
\end{figure}

In this section we present further analysis into the timing characteristics of real-world HPGP modems, as used in EVs and chargers.
As described in Section~\ref{sec:join_speed}, we analyzed captures collected by the SDR to study the delay between the encryption key being shared on the PLC network, to data packets being sent.
The time to first IPv6 packet sent by each of the three devices (EVs, chargers, and the MitM) is plotted in Figure~\ref{fig:join_speed}.
Additionally, we plot the time to the first SDP request sent by the vehicle.

\begin{figure*}[t]
    \centering
    \includegraphics[width=0.95\linewidth,trim={0cm 15cm 6.8cm 0cm},clip]{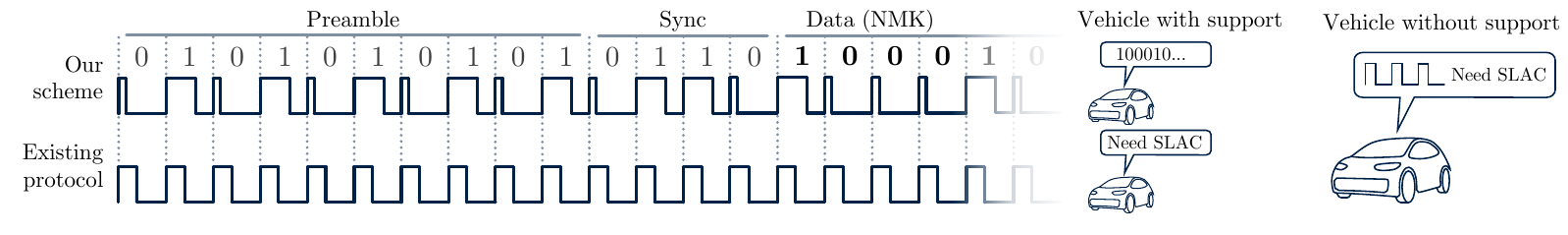}
    \caption{Illustration of the Control Pilot voltage using our new wireless attacker resistant scheme. The duty cycles have been exaggerated, the real protocol would use 4 and \SI{6}{\percent}. The vehicle with support recognizes the data stream, or defaults to SLAC without it. The vehicle without support cannot differentiate between our signal and the normal \SI{5}{\percent} carrier, and uses SLAC.}
    \Description{}
    \label{fig:new_proto}
\end{figure*}

We see that there is a large, consistent delay before EVs begin sending their SDP request packets.
We assume that this is because according to the ISO 15118-2 standard, the vehicle can only begin sending SDP requests after the modem has signaled that the HPGP network has been joined, and it is ready.
However, the EVs do send other IPv6 packets before this.
We found that these packets are various ICMPv6 messages, e.g. Router Solicitation or Neighbor Discovery.
It is unclear if they are sent immediately by the host system, or if they have already been queued inside the modem, waiting for a connection.
In any case, the plot shows that the SDR (\legendline{cyclergreen}{solid}) is faster than the EVs (\legendline{cyclerblue}{solid}), allowing us to always send packets before the EV has a chance to.

\section{NA Attack Details}
\label{app:na}

In this section we describe additional details of the NA Spoofing Attack.
The protocol diagram for this attack is shown in Figure~\ref{fig:na_attack}.
The goal is to rewrite the ND Cache within the charger, and thereby map the IP address of the EV to an attacker controlled MAC address.
To achieve this, the attacker needs to achieve two smaller challenges.

First, the attacker must know the IPv6 address of the EV, in order to create the correct mapping.
In our experiments, we found that all EVs calculate their IPv6 address based on their MAC address, using the EUI64 format.
This finding is also supported by a much larger study~\cite{szakaly2025current}, where they found that the majority of real-world chargers also do the same.
The MAC address of the EV is already used during the SLAC process, therefore the attacker has plenty of time to capture this, and predict the IP address.
Alternatively, as revealed in Appendix~\ref{app:join_speed}, it is common for EV modems to send IP packets long before they send their SDP request.
This means that they inadvertently leak their IP address to the attacker, enabling the NA spoofing attack.

Second, the attacker must ensure that the entry already exists within the victims ND cache, since the IPv6 standard allows NA packets to modify cache entries, but not create them.
A cache entry is only created when the victim system wishes to communicate with a specific destination.
At this point, it sends an ND request, creates the cache entry, and queues outbound packets until a response is received.
To force creation of this entry, we send a Ping packet with a spoofed source IP address, appearing to come from the EV.
Immediately while processing this, the software creates the entry, and queues the ping response.
Without needing to wait for the ND request, we can follow the ping with an NA packet, thereby populating the cache entry.

\end{document}